\documentclass[twocolumn,
pra
,superscriptaddress
,floatfix
,aps
,10pt
,showpacs]{revtex4-2}

\usepackage{graphicx}
\usepackage{amssymb}
\usepackage{amsmath}
\usepackage{multirow}
\usepackage{array,tabularx}
\usepackage{xcolor}
\usepackage{comment}

\usepackage[colorlinks,urlcolor=blue,citecolor=blue,linkcolor=blue]{hyperref}

\usepackage{textcomp}

\usepackage{float}
\usepackage{upgreek}
\usepackage{makeidx}
\DeclareGraphicsExtensions{.png,.jpg,.pdf,.mps,.gif,.bmp,.eps}
\graphicspath{{figures/}}
\usepackage{physics}

\begin{document}

\title{Unidirectional lasing in nonlinear Taiji micro-ring resonators}

\author{A. Mu$\tilde{\mathrm{n}}$oz de las Heras}
\email{a.munozdelasheras@unitn.it}
\affiliation{INO-CNR BEC Center and Dipartimento di Fisica, Universit$\grave{a}$ di Trento, 38123 Trento, Italy}

\author{I. Carusotto}
\affiliation{INO-CNR BEC Center and Dipartimento di Fisica, Universit$\grave{a}$ di Trento, 38123 Trento, Italy}

\date{\today}

\begin{abstract}

We develop a general formalism to study laser operation in active micro-ring resonators supporting two counterpropagating modes. Our formalism is based on the coupled-mode equations of motion for the field amplitudes in the two counterpropagating modes and a linearized analysis of the small perturbations around the steady state. We show that the devices including an additional S-shaped waveguide establishing an unidirectional coupling between both modes ---the so-called “Taiji” resonators (TJR)--- feature a preferred chirality on the laser emission and can ultimately lead to unidirectional lasing even in the presence of sizable backscattering. The efficiency of this mode selection process is further reinforced by the Kerr nonlinearity of the material. This stable unidirectional laser operation can be seen as an effective breaking of $\mathcal{T}$-reversal symmetry dynamically induced by the breaking of the $\mathcal{P}$-symmetry of the underlying device geometry. This mechanism appears as a promising building block to ensure non-reciprocal behaviors in integrated photonic networks and topological lasers without the need for magnetic elements.
\end{abstract}

\maketitle


\section{Introduction}
\label{sec:Introduction}

In the last decades silicon photonics platforms employing infrared light have become a prolific playground due to the well-established fabrication processes and the promising perspectives of integration in photonic circuits which are expected to form the core of next-generation information processing systems~\cite{Kaminow_2008,Nagarajan_2010}. 
In particular, ring resonator lasers have always been the subject of a great attention for both the interesting physics they host  and their rich technological applications~\cite{Chow_1985}. 

These systems support two degenerate counterpropagating modes in clockwise (CW) and counterclockwise (CCW) directions which are in strong mode competition. It has been shown that ring lasers can host stable unidirectional lasing in each direction~\cite{Burkhardt_2015}, yet with the undesired possibility of spontaneous switching between the two states due to quantum fluctuations~\cite{Singh_1979,Lett_1981,Zeghlache_1988,Mezosi_2009}. Other unsought features include backscattering processes coupling the two counterpropagating modes and even light localization at defects~\cite{Rawwagah_2006,Schwartz_2006,Schwartz_2007,Morthier_2013,Ceppe_2019}. This leads to a finite emission in the two directions and a spectral broadening of the emission with possible multi-mode behaviours and can even place the system in a self-oscillation regime where the intensity and phase of the emission periodically vary in time.

Preferential laser oscillation in one of the two counterpropagating modes is an appealing milestone due to the features that come alongside: increased output power, single-frequency spectrum, closer overlap with the gain profile, and improved mode stability, among others~\cite{Siegman_1986}. 
Stable unidirectional lasing in ring resonators has been investigated by adding an S-shaped waveguide element to the ring resonator, so to form a so-called ``Taiji'' resonator (TJR). This S-shaped element breaks spatial reflection $\mathcal{P}$-symmetry as it allows light of one mode to couple into the other one but forbids the opposite process. Robust unidirectional operation in this laser was first demonstrated in the near-infrared in~\cite{Hohimer_1993,Hohimer_1993b} and then has been further extended to other wavelengths and cavity designs~\cite{Shi_1995,Kharitonov_2015,Sacher_2015}.

The TJR has found further application in parity-time ($\mathcal{PT}$) symmetric micro-ring lasers~\cite{Ren_2018} and non-Hermitian active structures~\cite{Liu_2021}, and holds great promise to explore the physics of exceptional points~\cite{Soleymani_2021}. The idea of controlling the chirality of the laser emission via $\mathcal{P}$-symmetry breaking has also been exploited by establishing a loss imbalance~\cite{Lee_2016}, a coupling asymmetry between the two counterpropagating modes using two non-Hermitian nanoscatterers~\cite{Peng_2016}, and a bias in the pump direction~\cite{Cao_2020}. Very recently, a passive nonlinear TJR has been shown to break Lorentz reciprocity~\cite{MunozDeLasHeras_2021} and display direction-dependent transmission. As a recent new development, TJRs have started being employed~\cite{Harari_2018,Bandres_2018} as the building block of 2D topological lasers~\cite{Ozawa_2019,Ota_2020} in order to select a preferential chirality for the surface modes and prevent backscattering reflections.

Notwithstanding the proven importance of TJRs to achieve unidirectional lasing and their connections with deeply rooted physical ideas, we still lack a complete theory describing their operation. 
In this paper we develop a general theory of lasing in active resonators featuring sizable couplings between counter-propagating modes such as ring and Taiji resonators. We analyze the steady-state solutions of the coupled-mode equations of motion for the fields' amplitudes and we study their stability by looking at the small fluctuations dynamics. In this way we prove the crucial advantage of including an S-shaped element to guarantee robust unidirectional lasing and protect it against spurious backscattering-mediated mode couplings, e.g. by disorder in the resonator and interface roughness. Finally, we show how this protection is reinforced in nonlinear resonators displaying an intensity-dependent refractive index.

The Article is organized as follows: In Sec.~\ref{sec:SystemModel} we present the coupled-mode theory equations describing the system and the linearized approximation employed to assess the stability of their steady-state solutions. Sec.~\ref{sec:BackscatteringFreeResonators} makes use of these theoretical tools in order to study the laser emission of backscattering-free ring resonators (Sec.~\ref{sec:BSfreeRingResonator}) and TJRs (Sec.~\ref{sec:BSfreeTJR}). The effect of a weak backscattering is explored in Sec.~\ref{sec:MicBS}, where we show the robustness of unidirectional lasing in TJRs. This analysis is then extended to larger values of the  backscattering in Sec.~\ref{sec:LargeBackscattering}. The positive effect of the optical nonlinearity on the unidirectional lasing is highlighted in Sec.~\ref{sec:Nonlinear}. Conclusions are finally drawn in Sec.~\ref{sec:Conclusions}.


\section{The physical system and the theoretical model}
\label{sec:SystemModel}

\begin{figure}
    \centering
    \includegraphics[width=0.5\textwidth]{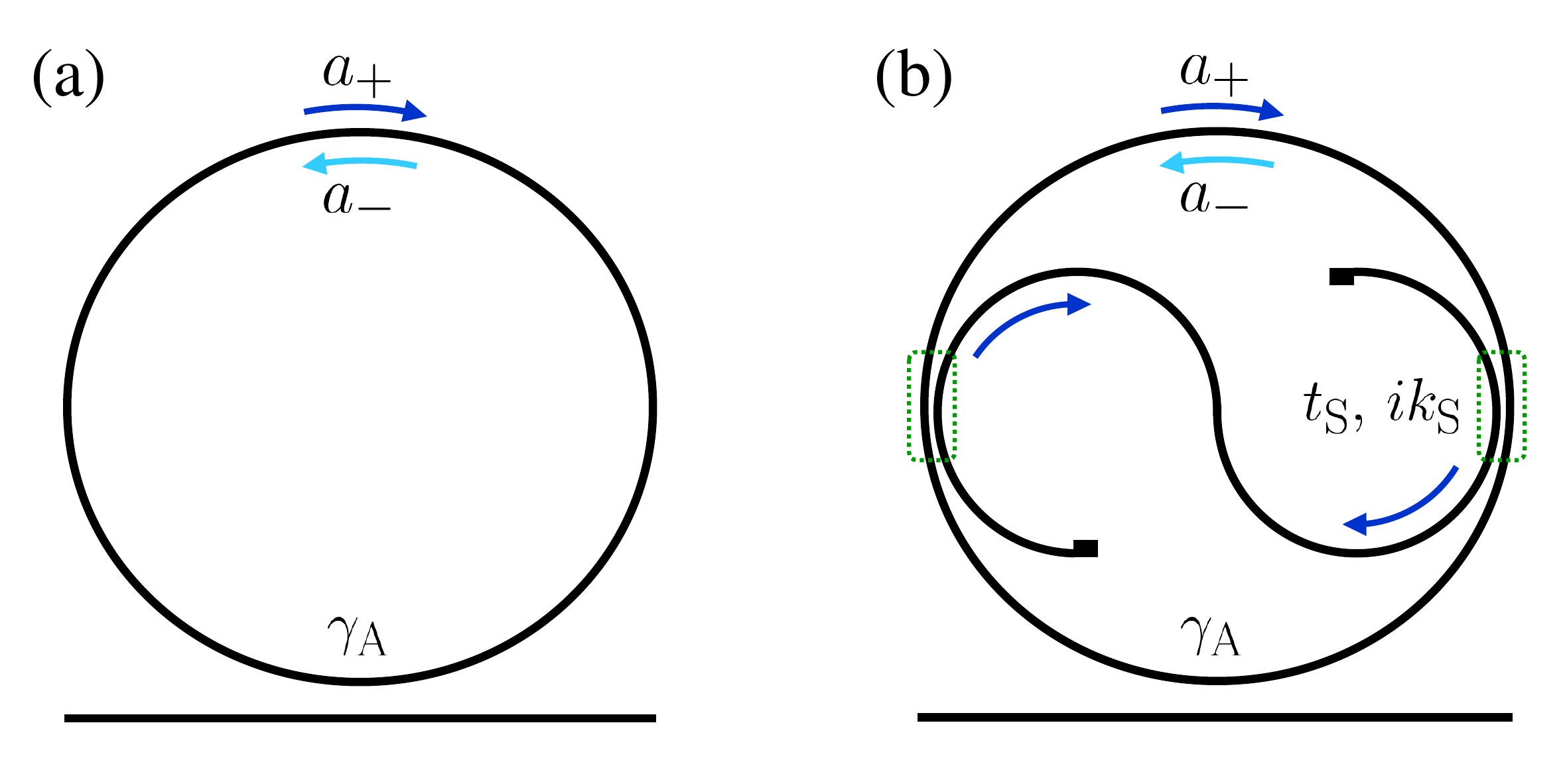}
    \caption{Schematic diagrams of the ring \textbf{(a)} and “Taiji” \textbf{(b)} microresonators. The field amplitudes of the clockwise (CW) and counterclockwise (CCW) modes are denoted by $a_{+}$ and $a_{-}$, respectively. In the Taiji microresonator (TJR) directional couplers of transmittance (coupling) amplitude $t_{\rm S}$ ($ik_{\rm S}$) are signalized with the dashed green rectangles. The loss rate $\gamma_{\rm A}$ accounts for absorption and radiative couplings, for instance with a bus waveguide.}
    \label{fig:TaijiDiagram}
\end{figure}

In this Section we mathematically describe an active resonator using the coupled-mode equations of motion for the field amplitudes of the CW and CCW modes. We then solve these equations for the long-time steady states for different values of the system parameters. We also present the linearized approximation that allows us to study the stability of the steady-states under small fluctuations. 

We consider a ring resonator of circumference $L_{0}$ (see Fig.~\ref{fig:TaijiDiagram}a) and the equivalent TJR with the same perimeter  enclosing an S-shaped waveguide (shown in Fig.~\ref{fig:TaijiDiagram}b). Without loss of generality the S waveguide was chosen to couple the CW into the CCW mode but not viceversa. The two elements of the TJR are coupled at two points separated by a distance $L_{\rm S}$ via lossless and reciprocal directional couplers with transmission and coupling amplitudes given by $t_{\rm S}$ and $ik_{\rm S}$, respectively ($t_{\rm S}$ and $k_{\rm S}$ are real numbers satisfying $t^2_{\rm S}+k^2_{\rm S}=1$). The S waveguide tips are assumed to feature a \textit{reflection killer} geometry where light is scattered away, thus preventing back-reflections~\cite{Castellan_2016}. The ring resonator case is recovered from the TJR equations by setting $t_{\rm S}=1$. Even though we considered a circular external perimeter, the model and its results are still valid for any other shape, like racetrack or square-like resonators. In realistic experiments the shape of the external waveguide can be optimized to reduce backscattering.

In all cases the resonator supports two counterpropagating modes in CW ($+$) and CCW ($-$) directions whose amplitudes $a_{\pm}$ can be described using the coupled-mode equations of motion~\cite{Henry_1982,WallsMilburn_1995,RevModPhys.85.299,Ghulinyan_2014,MunozDeLasHeras_2021}
\begin{align}
\label{eq:MotionEqsModes}
\dot{a}_{\pm} & =  -i\left[\omega_{0}a_{\pm} -\frac{n_{\text{NL}}}{n_{\text{L}}}\omega_{0}\left(|a_{\pm}|^2+g|a_{\mp}|^2\right)a_{\pm}
\right]\nonumber \\
&+\frac{P_{0}}{1+\frac{1}{n_{\text{S}}}(|a_{\pm}|^2+g|a_{\mp}|^2)}a_{\pm}-\gamma_{T}a_{\pm}\nonumber\\
&-i\beta_{\pm,\mp}a_{\mp}.
\end{align}
where $\omega_{0}$ is the resonance frequency of the resonator, $n_{\rm L}$ is the dimensionless linear refractive index, $n_{\rm NL}$ quantifies the strength of the optical nonlinearity, $g$ is a dimensionless parameter describing the character of the nonlinearity~\cite{MunozDeLasHeras_2021} ($g=1$ for a nonlocal thermo-optic nonlinearity and $g=2$ for a purely local Kerr-like one), $P_{0}$ is the amplification rate of the unsaturated gain, and $n_{\rm S}$ is the gain saturation coefficient. The total losses $\gamma_{\text{T}}=\gamma_{\rm A}+\gamma_{\rm S}$ include the intrinsic absorption and radiative losses $\gamma_{\rm A}$ of the ring (which for instance account for the coupling with a bus waveguide) and the effective losses due to radiation into the S waveguide $\gamma_{\rm S}=c\kappa^{2}_{\text{S}}/L_{0}n_{\rm L}$ (where $c$ is the vacuum speed of light and $\kappa_s$ is the amplitude for light insertion into the S waveguide). The generalized coupling coefficients $\beta_{\pm,\mp}$ account for all kinds of mode coupling, including backscattering and the S-waveguide couplings. A backscattering-free ring resonator features $\beta_{\pm,\mp}=0$, while our backscattering-free TJR is described by $\beta_{\pm}=0$, $\beta_{\mp}=-i2ck^{2}_{\rm S}e^{i\omega_{0} n_{\rm L}L_{\rm S}/c}/L_{0}n_{\rm L}$~\footnote{Of course all results in this paper are directly transferred to TJRs where the S waveguide recirculates light in the opposite direction from  the CCW mode into the CW one. To do this, it is enough to exchange the indices $+ \leftrightarrow -$.}.

In this coupled-mode formalism, the loss $\gamma_{\rm S,A,T}$ and amplification $P_0$ rates as well as the generalized coupling coefficients $\beta_{\pm,\mp}$ have the same units of inverse time as the resonance frequency $\omega_0$. On the other hand, some arbitrariness exists for the choice of the normalization and, even, the units of the field amplitudes $a_{\pm}$: in the quantum literature, they are often normalized so that their squared modulus $|a_{\pm}|^2$ give the number of photons present in each mode. In the photonics literature, one rather normalizes them so that their squared modulus $|a_{\pm}|^2$ corresponds to the light intensity (usually expressed in W$/$cm$^2$) circulating around the cavity in each mode. With this choice, $n_{\rm S}$ also has the dimension of an intensity, while $n_{\rm NL}$ has the usual meaning of a non-linear refractive index with the dimension of an inverse intensity and can be directly extracted from tables of material parameters.

Our calculations have been carried out with a silicon photonics implementation in mind. As typical parameters, we considered a resonator of perimeter $L_{0}=20$ $\mu$m and S-waveguide length $L_{\rm S}=L_{0}/2$, a refractive index $n_{\rm L}=3.5$ and a loss rate $\gamma_{\rm A}=6.5\times 10^{-6}$ $\mu$m$^{-1}=6.8\times 10^{9}$ s$^{-1}$.  For a resonance wavelength of $1.55$ $\mu$m, corresponding to a frequency $\omega_{0}\sim 2\pi\times 194$ THz, this  gives a Q-factor for the ring resonator (without S element) of about $Q\sim 1.5\times 10^{5}$.

But a key advantage of our coupled-mode formalism is its independence from the specific physical realization, where all physical parameters of the system are summarized into a few parameters. As one can see in what follows, all figures can be plotted in terms of adimensional quantities of transparent physical meaning: rates are normalized in units of the (experimentally accessible) loss rate or linewidth $\gamma_{\rm A,T}$ of the passive system; the amplification appears in the ratio $P_0/\gamma_{\rm T}$ characterizing the relative pumping compared to the laser threshold. These choices make the translation of our results into physical units a straightforward task for any given system once its basic parameters are known.

In order to simulate the response of the resonators we numerically solved Eq.~\ref{eq:MotionEqsModes} for the long-time ($t\rightarrow\infty$) steady states $a^{(0)}_{\pm}$ employing a 4th order Runge-Kutta algorithm and starting from random initial conditions at $t=0$. The random initial conditions serve as a seed of the dynamical instability of the trivial $a_{\pm}=0$ solution and trigger the onset of laser oscillation from the vacuum state.
As we will see in the following sections, in general the steady states oscillate at a single frequency and therefore it is useful to define $\tilde{a}_{\pm}=a_{\pm}e^{-i\omega t}$, where $\omega$ is a reference frequency. The stationary values of the field amplitudes $\tilde{a}^{(0)}_{\pm}=\tilde{a}_{\pm}(t\rightarrow\infty)$ satisfy $\partial_{t}\tilde{a}^{(0)}_{\pm}=0$ and therefore we can write the steady-state equation
\begin{align}
\label{eq:Ss}
0&=-i\left[(\omega_{0}-\omega)\tilde{a}^{(0)}_{\pm}
-\frac{n_{\text{NL}}}{n_{\text{L}}}\omega_{0}\left(|\tilde{a}^{(0)}_{\pm}|^2+g|\tilde{a}^{(0)}_{\mp}|^2\right)\tilde{a}^{(0)}_{\pm}\right] \nonumber \\
&+\frac{P_{0}}{1+\frac{1}{n_{\text{S}}}(|\tilde{a}^{(0)}_{\pm}|^2+g|\tilde{a}^{(0)}_{\mp}|^2)}\tilde{a}^{(0)}_{\pm}-\gamma_{\text{T}}\tilde{a}^{(0)}_{\pm}\nonumber\\
&-i\beta_{\pm,\mp}\tilde{a}^{(0)}_{\mp} \; .
\end{align}
%

We now describe the linearized approximation employed to assess the stability of the steady-state solutions to Eq.~\eqref{eq:MotionEqsModes}. We start by writing the field amplitudes $\tilde{a}_{\pm}$ as the sum of the stationary states $\tilde{a}^{(0)}_{\pm}$ satisfying Eq.~\eqref{eq:Ss} and the small fluctuations $|\delta\tilde{a}_{\pm}| \ll |\tilde{a}^{(0)}_{\pm}|$, i.e. $\tilde{a}_{\pm}=\tilde{a}^{(0)}_{\pm}+\delta\tilde{a}_{\pm}$. Introducing these expressions into Eq.~\eqref{eq:MotionEqsModes} and keeping fluctuations at linear order $\mathcal{O}(\delta\tilde{a}_{\pm})$ we arrive to the equations
\begin{align}
    \frac{d}{dt}
    \begin{bmatrix}
    \delta\tilde{a}_{+}&
    \delta\tilde{a}^{*}_{+}&
    \delta\tilde{a}_{-}&
    \delta\tilde{a}^{*}_{-}
    \end{bmatrix}^{T}
    =A
    \begin{bmatrix}
    \delta\tilde{a}_{+}&
    \delta\tilde{a}^{*}_{+}&
    \delta\tilde{a}_{-}&
    \delta\tilde{a}^{*}_{-}&
    \end{bmatrix}^{T},
\label{eq:FluctuationDynamics}
\end{align}
where the matrix $A$ is given by
%
%
\begin{widetext}
\begin{align}
\scriptsize
A=
\begin{bmatrix}
\begin{matrix}
-i(\omega_{0}-\omega)+2i\frac{n_{\text{NL}}}{n_{\text{L}}}\omega_{0}|\tilde{a}^{(0)}_{+}|^2\\
+i\frac{n_{\text{NL}}}{n_{\text{L}}}\omega_{0}g|\tilde{a}^{(0)}_{-}|^2\\
+\frac{P_{0}}{1+\frac{1}{n_{\text{S}}}(|\tilde{a}^{(0)}_{+}|^2+g|\tilde{a}^{(0)}_{-}|^2)}\\
-\gamma_{\text{T}}\\
-\frac{P_{0}/n_{\text{S}}}{\left[1+\frac{1}{n_{\text{S}}}(|\tilde{a}^{(0)}_{+}|^2+g|\tilde{a}^{(0)}_{-}|^2)\right]^2}\\
\times|\tilde{a}^{(0)}_{+}|^2
\end{matrix}
&
\begin{matrix}
i\frac{n_{\text{NL}}}{n_{\text{L}}}\omega_{0}\tilde{a}^{(0) 2}_{+}\\
-\frac{P_{0}/n_{\text{S}}}{\left[1+\frac{1}{n_{\text{S}}}(|\tilde{a}^{(0)}_{+}|^2+g|\tilde{a}^{(0)}_{-}|^2)\right]^2}\\
\times\tilde{a}^{(0) 2}_{+}
\end{matrix}
&
\begin{matrix}
i\frac{n_{\text{NL}}}{n_{\text{L}}}\omega_{0}g\tilde{a}^{(0) *}_{-}\tilde{a}^{(0)}_{+}\\
-i\beta_{\pm}\\
-\frac{P_{0}/n_{\text{S}}}{\left[1+\frac{1}{n_{\text{S}}}(|\tilde{a}^{(0)}_{+}|^2+g|\tilde{a}^{(0)}_{-}|^2)\right]^2}\\
\times g \tilde{a}^{(0) *}_{-}\tilde{a}^{(0)}_{+}
\end{matrix}
&
\begin{matrix}
i\frac{n_{\text{NL}}}{n_{\text{L}}}\omega_{0}g\tilde{a}^{(0)}_{-}\tilde{a}^{(0)}_{+}\\
-\frac{P_{0}/n_{\text{S}}}{\left[1+\frac{1}{n_{\text{S}}}(|\tilde{a}^{(0)}_{+}|^2+g|\tilde{a}^{(0)}_{-}|^2)\right]^2}\\
\times g\tilde{a}^{(0)}_{-}\tilde{a}^{(0)}_{+}
\end{matrix}
\\
\begin{matrix}
-i\frac{n_{\text{NL}}}{n_{\text{L}}}\omega_{0}\tilde{a}^{(0) * 2}_{+}\\
-\frac{P_{0}/n_{\text{S}}}{\left[1+\frac{1}{n_{\text{S}}}(|\tilde{a}^{(0)}_{+}|^2+g|\tilde{a}^{(0)}_{-}|^2)\right]^2}\\
\times\tilde{a}^{(0) * 2}_{+}
\end{matrix}
&
\begin{matrix}
i(\omega_{0}-\omega)-2i\frac{n_{\text{NL}}}{n_{\text{L}}}\omega_{0}|\tilde{a}^{(0)}_{+}|^2\\
-i\frac{n_{\text{NL}}}{n_{\text{L}}}\omega_{0}g|\tilde{a}^{(0)}_{-}|^2\\
+\frac{P_{0}}{1+\frac{1}{n_{\text{S}}}(|\tilde{a}^{(0)}_{+}|^2+g|\tilde{a}^{(0)}_{-}|^2)}\\
-\gamma_{\text{T}}\\
-\frac{P_{0}/n_{\text{S}}}{\left[1+\frac{1}{n_{\text{S}}}(|\tilde{a}^{(0)}_{+}|^2+g|\tilde{a}^{(0)}_{-}|^2)\right]^2}\\
\times|\tilde{a}^{(0)}_{+}|^2
\end{matrix}
&
\begin{matrix}
-i\frac{n_{\text{NL}}}{n_{\text{L}}}\omega_{0}g\tilde{a}^{(0) *}_{-}\tilde{a}^{(0) *}_{+}\\
-\frac{P_{0}/n_{\text{S}}}{\left[1+\frac{1}{n_{\text{S}}}(|\tilde{a}^{(0)}_{+}|^2+g|\tilde{a}^{(0)}_{-}|^2)\right]^2}\\
\times g\tilde{a}^{(0) *}_{-}\tilde{a}^{(0) *}_{+}
\end{matrix}
&
\begin{matrix}
-i\frac{n_{\text{NL}}}{n_{\text{L}}}\omega_{0}g\tilde{a}^{(0)}_{-}\tilde{a}^{(0) *}_{+}\\
+i\beta^{*}_{\pm}\\
-\frac{P_{0}/n_{\text{S}}}{\left[1+\frac{1}{n_{\text{S}}}(|\tilde{a}^{(0)}_{+}|^2+g|\tilde{a}^{(0)}_{-}|^2)\right]^2}\\
\times g\tilde{a}^{(0)}_{-}\tilde{a}^{(0) *}_{+}
\end{matrix}
\\
\begin{matrix}
i\frac{n_{\text{NL}}}{n_{\text{L}}}\omega_{0}g\tilde{a}^{(0) *}_{+}\tilde{a}^{(0)}_{-}\\
-i\beta_{\mp}\\
-\frac{P_{0}/n_{\text{S}}}{\left[1+\frac{1}{n_{\text{S}}}(|\tilde{a}^{(0)}_{-}|^2+g|\tilde{a}^{(0)}_{+}|^2)\right]}\\
\times g\tilde{a}^{(0) *}_{+}\tilde{a}^{(0)}_{-}
\end{matrix}
&
\begin{matrix}
i\frac{n_{\text{NL}}}{n_{L}}\omega_{0}g\tilde{a}^{(0)}_{+}\tilde{a}^{(0)}_{-}\\
-\frac{P_{0}/n_{\text{S}}}{\left[1+\frac{1}{n_{\text{S}}}(|\tilde{a}^{(0)}_{-}|^2+g|\tilde{a}^{(0)}_{+}|^2)\right]^2}\\
\times g\tilde{a}^{(0)}_{+}\tilde{a}^{(0)}_{-}
\end{matrix}
&
\begin{matrix}
-i(\omega_{0}-\omega)+2i\frac{n_{\text{NL}}}{n_{\text{L}}}\omega_{0}|\tilde{a}^{(0)}_{-}|^2\\
+i\frac{n_{\text{NL}}}{n_{\text{L}}}\omega_{0}g|\tilde{a}^{(0)}_{+}|^2\\
+\frac{P_{0}}{1+\frac{1}{n_{\text{S}}}(|\tilde{a}^{(0)}_{-}|^2+g|\tilde{a}^{(0)}_{+}|^2)}\\
-\gamma_{\text{T}}\\
-\frac{P_{0}/n_{\text{S}}}{\left[1+\frac{1}{n_{\text{S}}}(|\tilde{a}^{(0)}_{-}|^2+g|\tilde{a}^{(0)}_{+}|^2)\right]^2}\\
\times |\tilde{a}^{(0)}_{-}|^2
\end{matrix}
&
\begin{matrix}
i\frac{n_{\text{NL}}}{n_{\text{L}}}\omega_{0}\tilde{a}^{(0) 2}_{-}\\
-\frac{P_{0}/n_{\text{S}}}{\left[1+\frac{1}{n_{\text{S}}}(|\tilde{a}^{(0)}_{-}|^2+g|\tilde{a}^{(0)}_{+}|^2)\right]^2}\\
\times\tilde{a}^{(0) 2}_{-}
\end{matrix}
\\
\begin{matrix}
-i\frac{n_{\text{NL}}}{n_{\text{L}}}\omega_{0}g\tilde{a}^{(0) *}_{+}\tilde{a}^{(0) *}_{-}\\
-\frac{P_{0}/n_{\text{S}}}{\left[1+\frac{1}{n_{\text{S}}}(|\tilde{a}^{(0)}_{-}|^2+g|\tilde{a}^{(0)}_{+}|^2)\right]^2}\\
\times g\tilde{a}^{(0) *}_{+}\tilde{a}^{(0) *}_{-}
\end{matrix}
&
\begin{matrix}
-i\frac{n_{\text{NL}}}{n_{\text{L}}}\omega_{0}g\tilde{a}^{(0)}_{+}\tilde{a}^{(0) *}_{-}\\
+i\beta^{*}_{\mp}\\
-\frac{P_{0}/n_{\text{S}}}{\left[1+\frac{1}{n_{\text{S}}}(|\tilde{a}^{(0)}_{-}|^2+g|\tilde{a}^{(0)}_{+}|^2)\right]^2}\\
\times g\tilde{a}^{(0)}_{+}\tilde{a}^{(0) *}_{-}
\end{matrix}
&
\begin{matrix}
-i\frac{n_{\text{NL}}}{n_{\text{L}}}\omega_{0}\tilde{a}^{(0) * 2}_{-}\\
-\frac{P_{0}/n_{\text{S}}}{\left[1+\frac{1}{n_{\text{S}}}(|\tilde{a}^{(0)}_{-}|^2+g|\tilde{a}^{(0)}_{+}|^2)\right]^2}\\
\times \tilde{a}^{(0) * 2}_{-}
\end{matrix}
&
\begin{matrix}
i(\omega_{0}-\omega)-2i\frac{n_{\text{NL}}}{n_{\text{L}}}\omega_{0}|\tilde{a}^{(0)}_{-}|^2\\
-i\frac{n_{\text{NL}}}{n_\text{L}}\omega_{0}g|\tilde{a}^{(0)}_{+}|^2\\
+\frac{P_{0}}{1+\frac{1}{n_{\text{S}}}(|\tilde{a}^{(0)}_{-}|^2+g|\tilde{a}^{(0)}_{+}|^2)}\\
-\gamma_{\text{T}}\\
-\frac{P_{0}/n_{\text{S}}}{\left[1+\frac{1}{n_{\text{S}}}(|\tilde{a}^{(0)}_{-}|^2+g|\tilde{a}^{(0)}_{+}|^2)\right]^2}\\
\times |\tilde{a}^{(0)}_{-}|^2
\end{matrix}
\end{bmatrix}.
\label{eq:DiffEqsMatrix}
\end{align}
\end{widetext}
Altogether, Eqs.~(\ref{eq:FluctuationDynamics}-\ref{eq:DiffEqsMatrix}) describe the fluctuation dynamics. The steady states described by Eq.~\eqref{eq:Ss} are asymptotically stable whenever all eigenvalues of the matrix $A$ in Eq.~\eqref{eq:DiffEqsMatrix} have a negative real part. This means that fluctuations are damped out as $t\rightarrow\infty$ and the steady states behave as attractors in the $(a_{+},a_{-})$ space.

\begin{figure}
    \centering
    \includegraphics[width=0.5\textwidth]{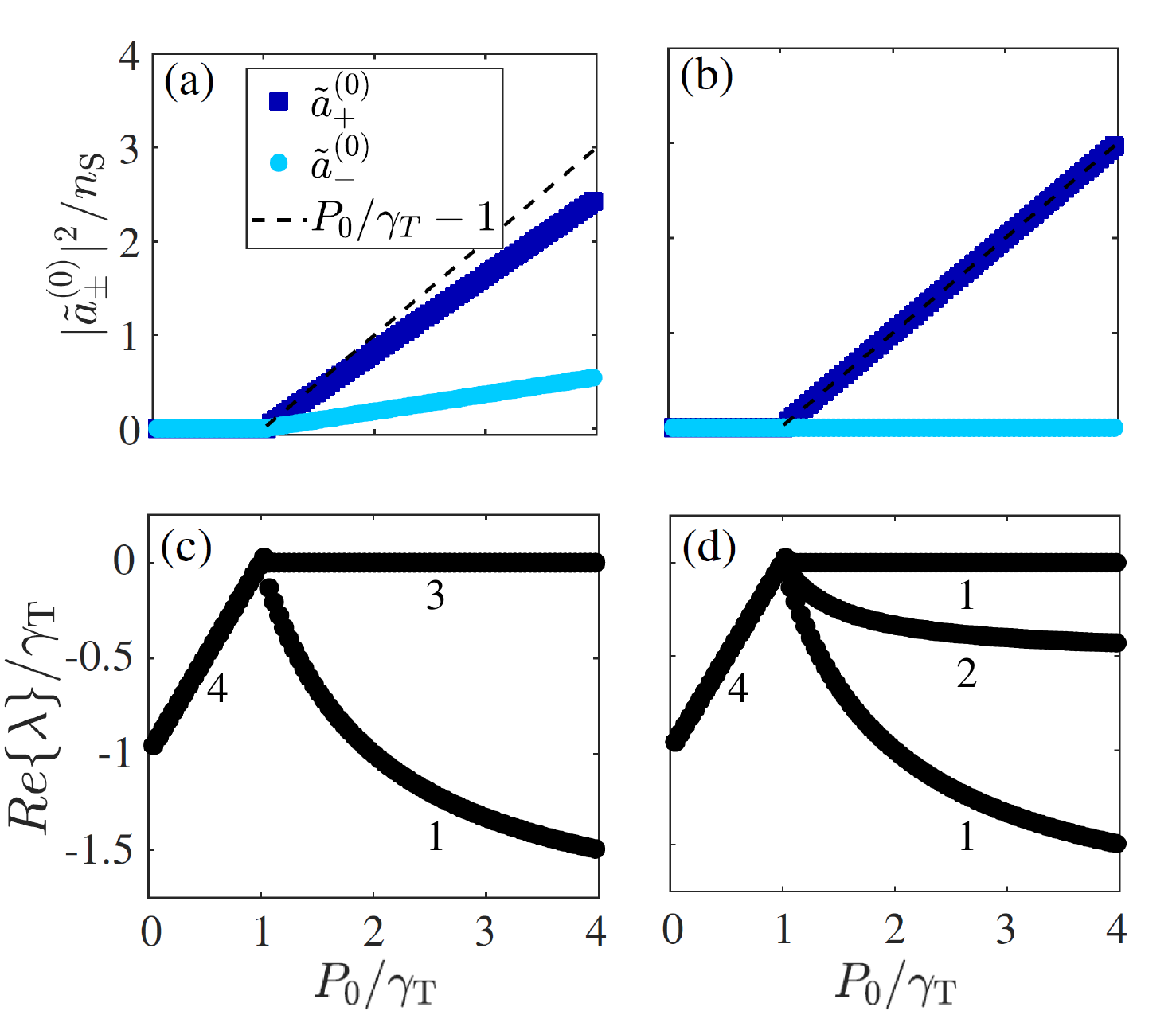}
    \caption{\textbf{(a,b)} Steady-state intensity of the CW ($|\tilde{a}^{(0)}_{+}|^2$) and CCW ($|\tilde{a}^{(0)}_{-}|^2$) modes for a backscattering-free ring resonator laser as a function of the pump rate $P_{0}$. Panel (a) refers to a nonlocal thermo-optic nonlinearity ($g=1$), while panel (b) corresponds to a local Kerr nonlinearity ($g=2$). The black dashed line represents the total intensity summed over the two directions.
    \textbf{(c,d)} Real part of the system eigenvalues $\lambda$ for a ring resonator laser  as a function of the pump rate $P_{0}$. Panel (c) refers to a nonlocal thermo-optic nonlinearity ($g=1$), while panel (d) corresponds to a local Kerr nonlinearity ($g=2$). The numbers below the curves indicate their degeneracy.}
    \label{fig:RingLasingStability}
\end{figure}
\begin{figure}
    \centering
    \includegraphics[width=0.5\textwidth]{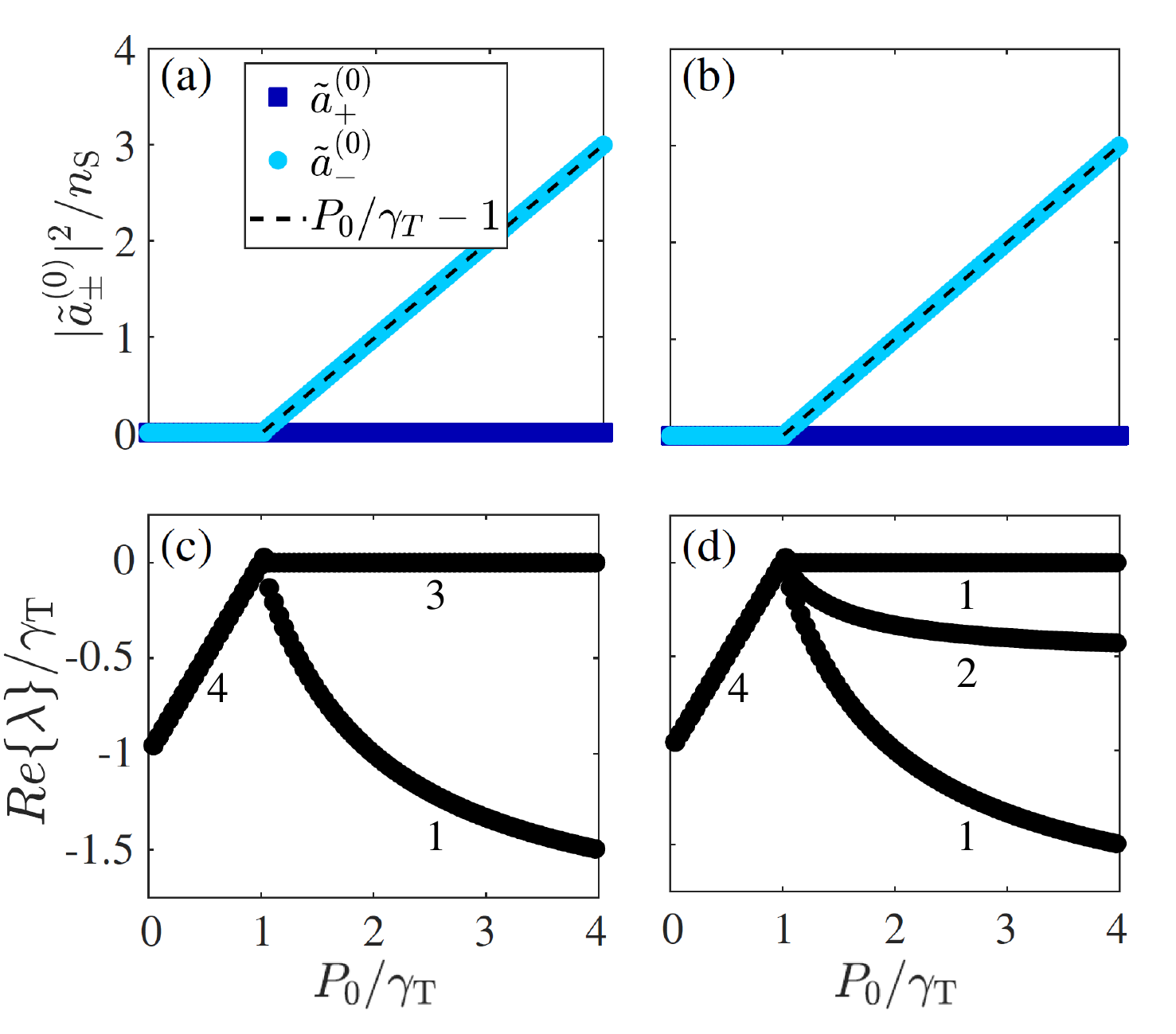}
    \caption{\textbf{(a,b)} Steady-state intensity of the CW ($|\tilde{a}^{(0)}_{+}|^2$) and CCW ($|\tilde{a}^{(0)}_{-}|^2$) modes for a backscattering-free TJR laser as a function of the pump rate $P_{0}$. Panel (a) refers to a nonlocal thermo-optic nonlinearity ($g=1$), while panel (b) corresponds to a local Kerr nonlinearity ($g=2$). The black dashed line represents the total intensity summed over the two directions.
    \textbf{(c,d)} Real part of the system eigenvalues $\lambda$ for a TJR laser  as a function of the pump rate $P_{0}$. Panel (c) refers to a nonlocal thermo-optic nonlinearity ($g=1$), while panel (d) corresponds to a local Kerr nonlinearity ($g=2$). The numbers below the curves indicate their degeneracy.}
    \label{fig:TaijiLasingStability}
\end{figure}
%


\section{Backscattering-free resonators}
\label{sec:BackscatteringFreeResonators}

In this Section we present the results for the paradigmatic cases of a ring resonator and a TJR without backscattering. We solve Eq.~\eqref{eq:MotionEqsModes} to find the possible steady-state solutions and plug them into Eqs.~(\ref{eq:FluctuationDynamics}-\ref{eq:DiffEqsMatrix}) in order to determine their stability. The analysis in this Section is valid in both the linear and nonlinear regimes. Without loss of generality, the field amplitudes $\tilde{a}_{\pm}$ can be taken as real quantities by choosing a rotating reference frame co-moving with the (possibly nonlinear-shifted) resonance frequency of the resonator. 

%
%


\subsection{Ring resonator}
\label{sec:BSfreeRingResonator}

We begin by studying the solutions to Eq.~\eqref{eq:MotionEqsModes} in the simplest backscattering-free ring resonator case (i.e. $t=1$, $\beta_{\pm,\mp}=0$). Fig.~\ref{fig:RingLasingStability} shows the steady-state intensities $|\tilde{a}^{0}_{\pm}|^2$ of both modes (panels (a) and (b)) together with the real parts of the eigenvalues of the fluctuation dynamics matrix~\eqref{eq:MotionEqsModes} (panels (c) and (d)) for growing values of the pump rate from $P_{0}=0$ to $P_{0}=4\gamma_{\rm T}$ in the $g=1$ (panels (a) and (c)) and $g=2$ (panels (b) and (d)) cases. In both situations, below the lasing threshold $P_{0}<\gamma_{\rm T}$ the intensity $|\tilde{a}^{(0)}_{\pm}|^2$ of both modes is zero and the four eigenvalues
\begin{align}
\label{eq:EigenvaluesBelowThreshold12}
    \lambda_{1,2}=+(\omega_{0}-\omega)+i(P_{0}-\gamma_{\rm T}), \\
    \lambda_{3,4}=-(\omega_{0}-\omega)+i(P_{0}-\gamma_{\rm T}),
\label{eq:EigenvaluesBelowThreshold34}
\end{align}
feature the same negative real part. Since gain saturates with $\tilde{a}^{(0)}_{\pm}$, in this regime the only possible stable solution is that no coherent light is present. Above the lasing threshold $P_{0}>\gamma_{\rm T}$ the eigenvalues change to
\begin{align}
\label{eq:EigenvaluesAboveThreshold1}
    \lambda_{1}/\gamma_{\rm T}&=-\frac{n_{\rm NL}}{n_{\rm L}}\omega_{0}(g-1)n_{\rm S}\left(\frac{P_{0}}{\gamma_{\rm T}}-1\right)\nonumber\\
    &+i\left(\frac{P_{0}/\gamma_{\rm T}}{1+g(P_{0}/\gamma_{\rm T}-1)}-1\right), \\
    \lambda_{2}/\gamma_{\rm T}&=+\frac{n_{\rm NL}}{n_{\rm L}}\omega_{0}(g-1)n_{\rm S}\left(\frac{P_{0}}{\gamma_{\rm T}}-1\right)\nonumber\\
    &+i\left(\frac{P_{0}/\gamma_{\rm T}}{1+g(P_{0}/\gamma_{\rm T}-1)}-1\right), \\
    \lambda_{3}&=0, \\
    \lambda_{4}/\gamma_{\rm T}&=-2i\frac{P_{0}/\gamma_{\rm T}-1}{P_{0}/\gamma_{\rm T}}.
\label{eq:EigenvaluesAboveThreshold4}
\end{align}

The $g=1$ case presents a particular behaviour as it is the only situation in which the system shows effective gain in both directions and features a threefold degenerate Goldstone mode with $\lambda_{1,2,3}=0$. This is due to the phase freedom of each direction and the CW $\leftrightarrow$ CCW symmetry which is present in this case only. Beyond the bifurcation point both intensities depart linearly with randomly-chosen values satisfying $(|\tilde{a}^{(0)}_{+}|^2+|\tilde{a}^{(0)}_{-}|^2)/n_{\rm S}=P_{0}/\gamma_{\rm T}-1$.

For any other value $1 < g \leq 2$ the resonator lases randomly in one mode only with equal probability as $|\tilde{a}^{(0)}_{\pm}|^2/n_{\rm S}=P_{0}/\gamma_{\rm T}-1$, while the other one remains unamplified. In contrast to the $g=1$ case, now the linearized analysis features a single Goldstone mode ($\lambda_{3}=0$, $\lambda_{1,2,4}\neq 0$). Due to the asymmetric gain amplification terms in Eq.~\eqref{eq:MotionEqsModes} a larger intensity in a certain direction randomly determined by the initial conditions  
leads to a smaller amplification in the other one. As $t$ evolves this asymmetry drives the system into a unidirectional lasing state in the randomly favored mode, while the other is killed.

By employing a Fourier transform we were able to ascertain that for $P_{0}>\gamma_{\rm T}$ the steady-state field amplitudes $a^{(0)}_{\pm}$ oscillate in time at the nonlinear-shifted resonance frequency of the resonator, which we take as the reference frequency $\omega$. In the $g=1$ case this is given by 
\begin{align}
    \omega=\omega_{0}-\frac{n_{\rm NL}}{n_{\rm L}}\omega_{0}(|\tilde{a}^{(0)}_{+}|^2+|\tilde{a}^{(0)}_{-}|^2),
    \label{eq:omega_g1}
\end{align}
while for $g>1$ the intensity in one of the two directions vanishes and one has either
\begin{align}
    \omega=\omega_{0}-\frac{n_{\rm NL}}{n_{\rm L}}\omega_{0}|\tilde{a}^{(0)}_{\pm}|^2
    \label{eq:omega_g2}
\end{align}
for unidirectional lasing in the CW or CCW direction, respectively.

The fact that no bidirectional lasing can be observed for $g>1$ can be put on more solid grounds as follows. From Eq.~\eqref{eq:Ss} one sees that if both field amplitudes $\tilde{a}^{(0)}_{\pm}$ are taken as real numbers a finite intensity in the two directions (i.e. $\tilde{a}^{(0)}_{\pm}\neq 0$) would simultaneously imply
\begin{align}
\label{eq:FreqLasingBothModes}
    \frac{P_{0}}{1+\frac{1}{n_{\rm S}}(|\tilde{a}^{(0)}_{\pm}|^2+g|\tilde{a}^{(0)}_{\mp}|^2)}=\gamma_{\rm T},
\end{align}
which necessarily sets $|\tilde{a}^{(0)}_{+}|^2=|\tilde{a}^{(0)}_{-}|^2$. Diagonalizing the matrix in this case, one obtains the eigenvalues
\begin{align}
    \lambda_{1,2}&=0,\\
    \lambda_{3}/\gamma_{\rm T}&=+2i\frac{P_{0}/\gamma_{\rm T}-1}{P_{0}/\gamma_{\rm T}}\frac{g-1}{g+1},\\
    \lambda_{4}/\gamma_{\rm T}&=-2i\frac{P_{0}/\gamma_{\rm T}-1}{P_{0}/\gamma_{\rm T}}.
\end{align}
Having $g>1$ implies $\Im{\lambda_{3}}>0$ and therefore this solution is unstable. We conclude that for $g>1$ the only possible stable solutions involve unidirectional lasing in one of the two counter-propagating modes, randomly chosen by the initial conditions.


\subsection{TJR}
\label{sec:BSfreeTJR}

We now move on to the study of the stability of an active TJR under the same ramp in the pump rate employed in the previous Subsection. For these simulations we set $\gamma_{\rm S}=0.2\gamma_{\rm A}$, $\beta_{\pm}=0$, and $\beta_{\mp}=-i2ck^{2}_{\rm S}e^{i\omega_{0} n_{\rm L}L_{\rm S}/c}/L_{0}n_{\rm L}$. 
We have neglected the nonlinear phase shift inside the S waveguide as the optical power inside it is very small compared to that circulating along the external waveguide. 

Fig.~\ref{fig:TaijiLasingStability} shows the steady-state intensities $|\tilde{a}^{(0)}_{\pm}|^2$ (in panels (a) and (b) for $g=1$ and $g=2$, respectively) and the real part of the eigenvalues (in panels (c) and (d) for $g=1$ and $g=2$, respectively) for the family of steady-state solutions that are most commonly reached. Below the lasing threshold $P_{0}<\gamma_{\rm T}$ the two directions feature a zero intensity for both interaction types. Above the lasing threshold $P_{0}>\gamma_{\rm T}$ the intensity of the mode in which the S-waveguide coupling is directed (in our case the CCW) departs linearly from the bifurcation as $|\tilde{a}^{(0)}_{-}|^2/n_{\rm S}=P_{0}/\gamma_{\rm T}-1$, while the other mode (the CW one) remains empty. The off-diagonal S-waveguide coupling in Eq.~\eqref{eq:DiffEqsMatrix} does not change the eigenvalues of the linearized theory w.r.t. the ring resonator case and therefore they are given again by Eqs.~(\ref{eq:EigenvaluesBelowThreshold12}-\ref{eq:EigenvaluesBelowThreshold34}) below threshold and by Eqs.~(\ref{eq:EigenvaluesAboveThreshold1}-\ref{eq:EigenvaluesAboveThreshold4}) above threshold. As opposed to the ring resonator case, not even for $g=1$ one can have a finite intensity in both modes simultaneously: the presence of the S waveguide breaks in fact the CW $\leftrightarrow$ CCW symmetry and favors unidirectional lasing in the CCW direction. As in the previous Subsection using a Fourier transform we found that above $P_{0}>\gamma_{\rm T}$ the steady-state field amplitudes $a^{(0)}_{\pm}$ oscillate at the nonlinear-shifted resonance frequency of the resonator, which is taken as the reference frequency $\omega$. For our TJR this is given by
\begin{align}
    \omega=\omega_{0}-\frac{n_{\rm NL}}{n_{\rm L}}\omega_{0}|\tilde{a}^{(0)}_{-}|^2
    \label{eq:omega_TJR}
\end{align}
for any value of $g$.

\begin{figure*}[t]
    \centering
    \includegraphics[width=\textwidth]{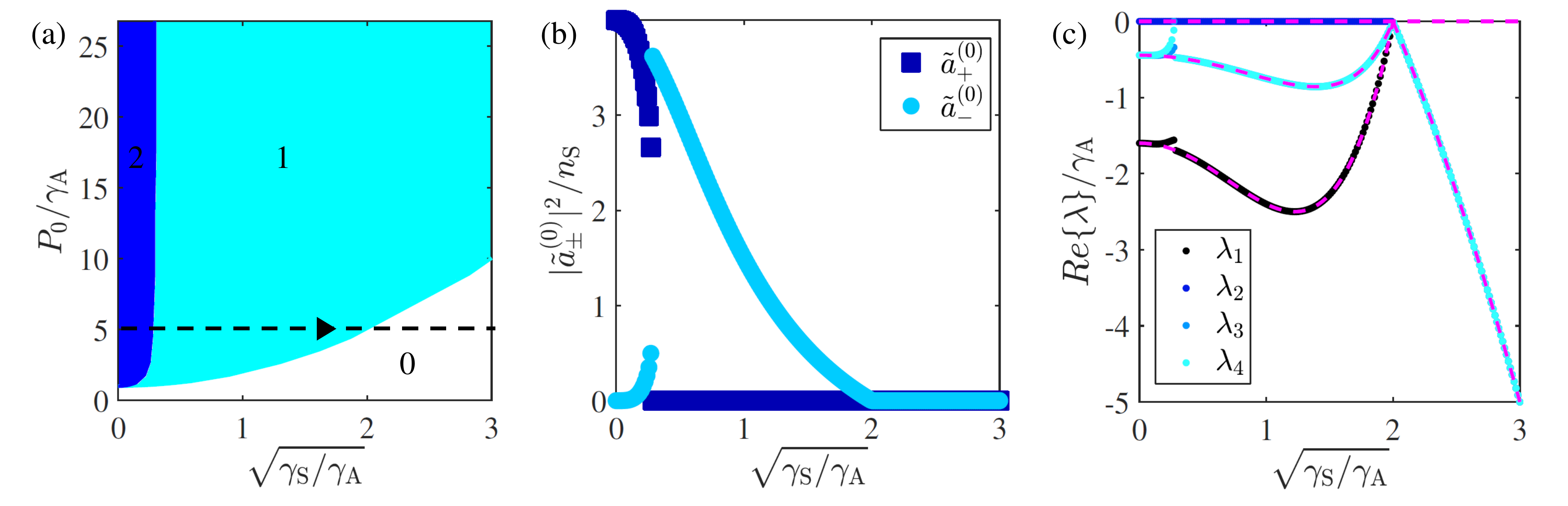}
    \caption{\textbf{(a)} Pump rate $P_{0}$ vs. square root of the S-coupling losses $\sqrt{\gamma_{\rm S}}$ diagram for a backscattering-free TJR with $g=2$ and $n_{\rm NL}=0$. The region where only the trivial solution is present is depicted in white. The light blue region represents the parameter range where only CCW lasing is possible. In the dark blue region two solutions exist: single-mode lasing in the preferred CCW direction and bidirectional lasing with $|\tilde{a}^{(0)}_{+}|\gg|\tilde{a}^{(0)}_{-}|$. Each region is labeled by the numbers 0, 1, and 2, respectively. \textbf{(b)} Normalized steady-state intensity $|\tilde{a}^{(0)}_{\pm}|^2$ obtained by numerically solving Eq.~\eqref{eq:MotionEqsModes} as a function of $\sqrt{\gamma_{\rm S}}$ for a fixed $P_{0}=5\gamma_{\rm A}$ as indicated by the dashed horizontal line in (a). \textbf{(c)} Real part of the system eigenvalues $\lambda$ as a function of $\sqrt{\gamma_{\rm S}}$ corresponding to the path shown as the dashed horizontal line of panel (a). The points are numerically calculated by diagonalizing the matrix $A$ in Eq.~\eqref{eq:DiffEqsMatrix} for the steady-state solutions displayed in panel (b). Where not visible, $Re\{\lambda_{3}\}$ lies below $Re\{\lambda_{4}\}$. Above $\sqrt{\gamma_{\rm S}/\gamma_{\rm A}}=2$ all real parts are degenerate. The dashed lines correpond to the analytic eigenvalues~(\ref{eq:EigenvaluesBelowThreshold12}-\ref{eq:EigenvaluesAboveThreshold4}).}
    \label{fig:SolutionsPhaseDiagram}
\end{figure*}

While the unidirectional CCW-only solutions depicted in Fig.~\ref{fig:TaijiLasingStability} are the steady-state that is most frequently reached by the numerics, it is important to verify whether other solutions are possible in specific parameter regimes. To this purpose, we considered  Eq.~\eqref{eq:Ss} assuming a finite intensity in both modes, i.e. $\tilde{a}^{(0)}_{\pm}\neq 0$. Without loss of generality $\tilde{a}^{(0)}_{+}$ is taken to be a real number. From Eq.~\eqref{eq:Ss} this would require
\begin{widetext}
\begin{align}
    \omega_{0}-\omega=\frac{n_{\rm NL}}{n_{\rm L}}\omega_{0}(|\tilde{a}^{(0)}_{+}|^2+g|\tilde{a}^{(0)}_{-}|^2),\\
    \label{eq:BSfreeTJR_OtherSolutions1}
    \frac{P_{0}}{1+\frac{1}{n_{\rm S}}(|\tilde{a}^{(0)}_{+}|^2+g|\tilde{a}^{(0)}_{-}|^2)}=\gamma_{\rm T},\\
    \label{eq:BSfreeTJR_OtherSolutions2}
    \tilde{a}^{(0)}_{-}=\frac{i\frac{c}{L_{0}n_{\text{L}}}2\kappa^2_{\text{S}}e^{i\omega_{0} n_{\rm L}L_{\rm S}/c}}{(\omega_{0}-\omega)-\frac{n_{\text{NL}}}{n_{\text{L}}}\omega_{0}\left(|\tilde{a}^{(0)}_{-}|^2+g|\tilde{a}^{(0)}_{+}|^2\right)+i\frac{P_{0}}{1+\frac{1}{n_{S}}(|\tilde{a}^{(0)}_{-}|^2+g|\tilde{a}^{(0)}_{+}|^2)}-i\gamma_{T}}\tilde{a}^{(0)}_{+}.
\end{align}
\end{widetext}
In the $g=1$ case the first two conditions impose the denominator in Eq.~\eqref{eq:BSfreeTJR_OtherSolutions2} to vanish, which implies that $\tilde{a}^{(0)}_{+}=0$.

To carry on this analysis in the $g>1$ case we explored the phase space of the problem by numerically solving Eq.~\eqref{eq:MotionEqsModes} using different values of $P_{0}$ and $\gamma_{\rm S}$ in order to explore its possible steady-state solutions. We found that stable bidirectional lasing is possible for values of $P_{0}$ and $\gamma_{\rm S}$ within the dark blue region of Fig.~\ref{fig:SolutionsPhaseDiagram}a. This kind of solutions involve a strong lasing in the CW direction and a smaller intensity populating the CCW due to the presence of a weak S coupling. For larger values of $\gamma_{\rm S}$, $|\tilde{a}^{(0)}_{-}|^2$ increases up to a threshold in which the $\tilde{a}^{(0)}_{+}\neq 0$ solution disappears and only the $\tilde{a}^{(0)}_{-}\neq 0$ one remains. While this figure is plotted in the linear ($n_{\rm NL}=0$) regime, we have verified that the same conclusion holds in the presence of nonlinearities, $n_{\rm NL}\neq 0$. However, the stronger is the nonlinearity $n_{\rm NL}$, the smaller is $|\tilde{a}^{(0)}_{-}|^2$ in the bidirectional solution. 

The light blue parameter range of Fig.~\ref{fig:RingLasingStability} features a single solution with emission in the CCW mode only. The lower limit is given by the pump threshold $P_{0}=\gamma_{\rm T}$. 
Of course, this CCW-mode lasing solution also extends in the dark-blue region of Fig.~\ref{fig:RingLasingStability} where it coexist with the other, bidirectional solution. Which of the two solutions is actually chosen by the system will depend on the initial conditions.

Panels (b) and (c) show the lasing intensity and the real part of the eigenvalues (calculated by diagonalizing the matrix $A$ in Eq.~\eqref{eq:DiffEqsMatrix}) for the ramp depicted as the dashed line in panel (a), respectively. The initial conditions correspond to unidirectional lasing in the CW direction. As $\gamma_{\rm S}$ is increased the initially unfavored CCW mode gets rapidly populated and the emission intensity in the CW mode decreases. As this happens the real part of one of the eigenvalues grows and eventually crosses zero towards positive values. At this point the solution becomes unstable and the system experiences a transition towards a unidirectional lasing regime in the CCW direction. Note that as $\gamma_{\rm S}$ grows, the total loss rate $\gamma_{\rm T}$ is also enhanced so that the CCW intensity decreases until the system ends up in the trivial solution.

We conclude that the presence of the S-shaped element breaking $\mathcal{P}$-symmetry in Eq.~\eqref{eq:MotionEqsModes} rules out the possibility of pure unidirectional lasing solutions in the unfavored direction. For $g=1$ not even bidirectional emission is possible and beyond the lasing threshold the system compulsory lases in a single direction with the preferential chirality introduced by the S waveguide. For $g>1$ bidirectional lasing is possible for a small enough S-coupling, but only for particular values of the parameters and always giving $|\tilde{a}^{(0)}_{+}|^2 \gg |\tilde{a}^{(0)}_{-}|^2$.

From a more abstract perspective, we can conclude this section by noting that, for any value of $g$, in contrast to what happens in a passive device or for a pump rate $P_{0}<\gamma_{\rm T}$, the solution for $P_{0}>\gamma_{\rm T}$ is not invariant under $\mathcal{T}$-reversal even though the underlying equations of motion are fully $\mathcal{T}$-reversal symmetric. We can therefore state that the presence of the $\mathcal{P}$-breaking S-shaped element induces a dynamical breaking of $\mathcal{T}$-symmetry above the lasing threshold.


\section{Small backscattering}
\label{sec:MicBS}

In this Section we study how the laser emission of active ring resonators and TJRs is affected by a random backscattering smaller in modulus than the intrinsic absorption and radiative loss rate $\gamma_{\rm A}$ of the ring. The analog analysis for a backscattering larger than $\gamma_{\rm A}$ can be found in Sec.~\ref{sec:LargeBackscattering}. In both Sec.~\ref{sec:MicBS} and Sec.~\ref{sec:LargeBackscattering} the nonlinear shift of the resonance frequency of the resonator is neglected (i.e. we set $n_{\rm NL}=0$). This effect will be studied in Sec.~\ref{sec:Nonlinear}. As the pure nonlocal $g=1$ case is not representative of a realistic experiment, starting from the present Section we will focus exclusively in values within the range $1<g\leq 2$. As in Sec.~\ref{sec:BackscatteringFreeResonators} we employ the reference frequencies~\eqref{eq:omega_g2} and~\eqref{eq:omega_TJR} which for $n_{\rm NL}=0$ reduce to the resonance frequency of the resonator $\omega=\omega_{0}$.

In general, the effect of backscattering can be summarized by the Hermitian and non-Hermitian coefficients
\begin{align}
\label{eq:n}
    h&=-\frac{\beta_{\mp}+\beta^{*}_{\pm}}{2},\\
    n&=i\frac{\beta_{\mp}-\beta^{*}_{\pm}}{2},
\label{eq:h}
\end{align}
defined in~\cite{Biasi_2019}. The former gives rise to a symmetric, conservative exchange of energy between the CW and CCW modes, while the latter introduces a different balance between the back-reflection in each mode that can lead to gain and losses beyond the saturable gain ($\propto P_{0}$) and losses ($\propto \gamma_{\rm T}$) terms in Eq.~\eqref{eq:MotionEqsModes}. Therefore we need $|n|\leq\gamma_{\rm T}$ in order to preserve the validity of our model. 

The distinction between \textit{small} and \textit{large} backscattering made through this Article is referred to the loss rate without S couplings $\gamma_{\rm A}$. By small backscattering we intend that it is at least one order of magnitude smaller than $\gamma_{\rm A}$, i.e. $|h|,|n|\lesssim 10^{-1}\times\gamma_{\rm A}$. On the other hand, we labeled as large backscattering that which is at least comparable with $\gamma_{\rm A}$, i.e. $|h|\gtrsim 10^{-1}\gamma_{\rm A}$. For all parameter choices, simulations are carried by solving Eq.~\eqref{eq:MotionEqsModes} for different values of $n$ and $h$ and finding the $t\rightarrow\infty$ steady states $\tilde{a}^{(0)}_{\pm}$. As usual, the stability of the solutions is assessed by calculating the corresponding eigenvalues of the fluctuation dynamics matrix in Eq.~\eqref{eq:DiffEqsMatrix}.

\begin{figure}
    \centering
    \includegraphics[width=0.5\textwidth]{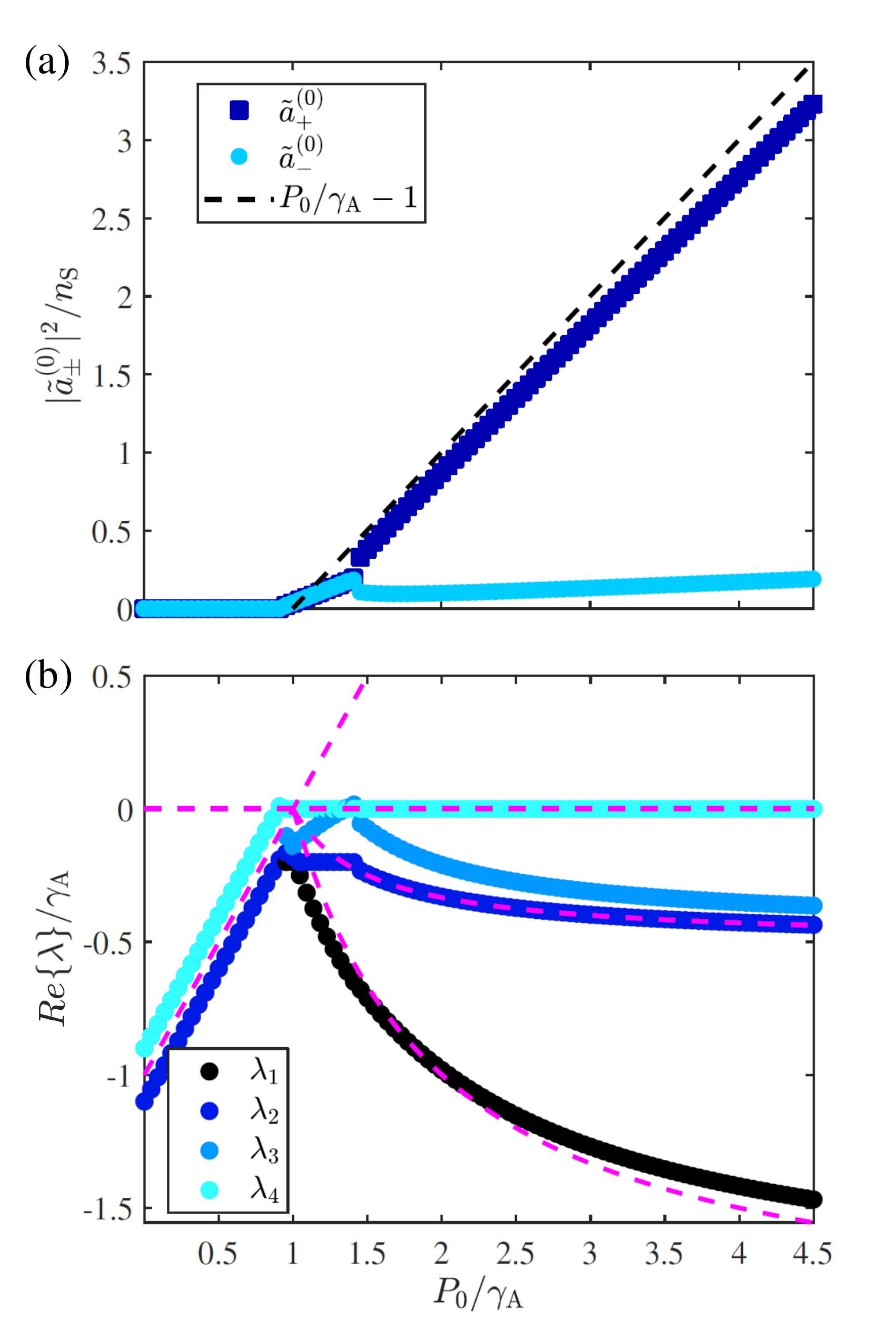}
    \caption{Ring resonator with $g=2$ and backscattering parameters $|n|=0.1\gamma_{\rm A}$ and $h=0$. \textbf{(a)} Steady-state intensity $|\tilde{a}^{(0)}_{\pm}|^2$ in each mode as a function of the pump rate $P_{0}$. The dashed line is the backscattering-free intensity. \textbf{(b)} Real part of the fluctuation dynamics eigenvalues $\lambda$ as a function of $P_{0}$. The dashed lines represent the real part of the backscattering-free eigenvalues~(\ref{eq:EigenvaluesBelowThreshold12}-\ref{eq:EigenvaluesAboveThreshold4}). Below threshold, $Re\{\lambda_{1}\}=Re\{\lambda_{2}\}$ and $Re\{\lambda_{3}\}=Re\{\lambda_{4}\}$.}
    \label{fig:SmallBS}
\end{figure}
%


\subsection{Perturbative solution}
\label{sec:PerturbativeSolution}

\begin{figure*}
    \centering
    \includegraphics[width=\textwidth]{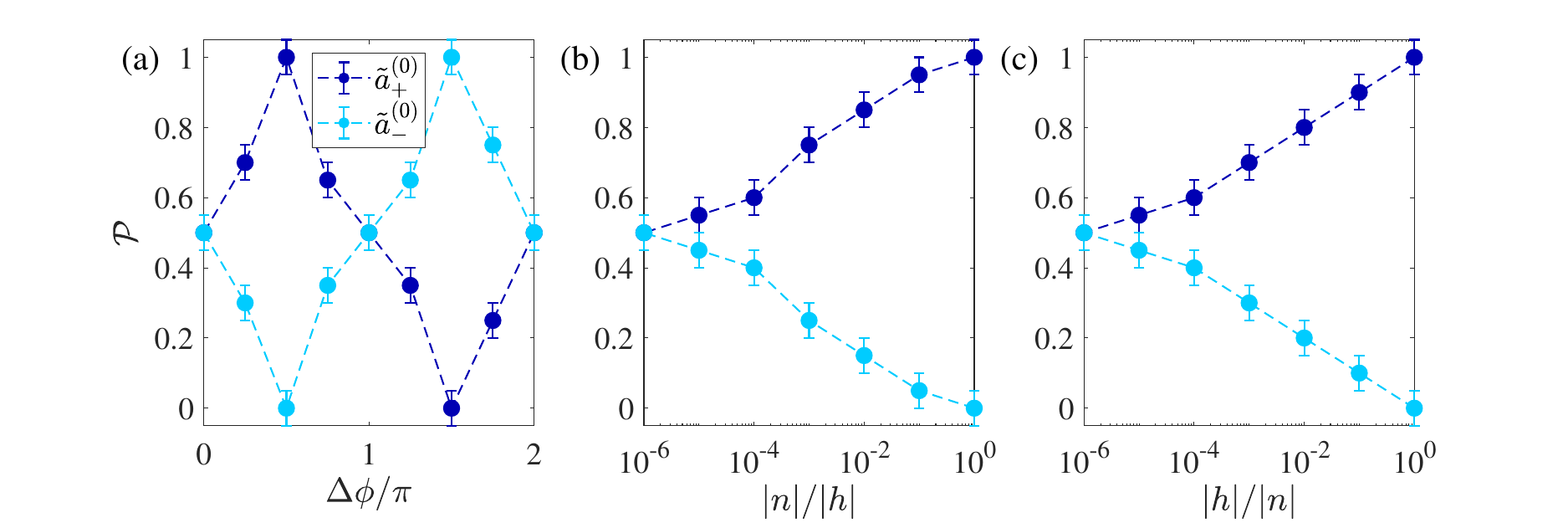}
    \caption{Probability $\mathcal{P}$ of a ring resonator with $g>1$ to lase preferentially in the CW (dark blue) and CCW (light blue) direction for different values of the backscattering. \textbf{(a)} Fixed modulus $|h|=|n| \simeq 5\times 10^{-4}\gamma_{\rm A}$ as a function of the relative phase angle $\Delta\phi=\phi_{n}-\phi_{h}$. \textbf{(b)} Fixed $\Delta\phi=\pi/2$ and $|h|\simeq 5\times 10^{-4}\gamma_{\rm A}$ as a function of $|n|$. \textbf{(c)} Fixed $\Delta\phi=\pi/2$ and $|n|\simeq 5\times 10^{-4}\gamma_{\rm A}$ as a function of $|h|$.}
    \label{fig:BackscatteringRing}
\end{figure*}

Prior to the numerics we discuss the perturbative analytical solution of Eq.~\eqref{eq:MotionEqsModes} valid for both the ring resonator and the TJR as microscopic backscattering is added. We assume that this is so small that the backscattering-free solution of Eq.~\eqref{eq:MotionEqsModes} involving unidirectional lasing does not substantially change. In order to account for the small intensity in the suppressed mode we considered the backscattering couplings $\tilde{\beta}_{\pm,\mp}$ as the small perturbations to the backscattering-free ring resonator and TJR  solutions explored in Sec.~\ref{sec:BackscatteringFreeResonators}. The new coupling parameters for the ring resonator are $\beta_{\pm,\mp}=\tilde{\beta}_{\pm,\mp}$, while those for the TJR are  $\beta_{\pm}=\tilde{\beta}_{\pm}$ and $\beta_{\mp}=-i2ck^{2}_{\rm S}e^{i\omega_{0} n_{\rm L}L_{\rm S}/c}/L_{0}n_{\rm L}+\tilde{\beta}_{\mp}$. Without loss of generality, and in order to use the same indices in both cases, we considered an unperturbed solution in which the ring resonator lases unidirectionally in the CCW mode. We can then write the perturbed solutions as $\tilde{a}^{(0)}_{\pm}=\tilde{a}^{(0)}_{\pm,\rm un}+\delta\tilde{a}^{(0)}_{\pm}$, where $\tilde{a}^{(0)}_{\pm,\rm un}$ are the unperturbed solutions with $\tilde{\beta}_{\pm,\mp}=0$. As shown in Sec.~\ref{sec:BackscatteringFreeResonators} these are given by $\tilde{a}^{(0)}_{+,\rm un}=0$ and $|\tilde{a}^{(0)}_{-,\rm un}|^2/n_{\rm S}=P_{0}/\gamma_{\rm T}-1$. On the other hand $\delta\tilde{a}^{(0)}_{\pm}$ are the first-order perturbative corrections. Introducing the perturbed solution into the steady-state given by Eq.~\eqref{eq:Ss} and staying at linear order $\mathcal{O}(\tilde{\beta}_{\pm,\mp})$, $\mathcal{O}(\delta\tilde{a}^{(0)}_{\pm})$ in the perturbation we arrive at
\begin{align}
\label{eq:PertTheory}
    \delta\tilde{a}^{(0)}_{+}&=\frac{i\tilde{\beta}_{\pm}}{\frac{P_{0}}{1+\frac{g}{n_{\rm S}}|\tilde{a}^{(0)}_{-,\rm un}|^2}-\gamma_{\rm T}}\tilde{a}^{(0)}_{-,\rm un}.
\end{align}
The intensity in the perturbatively populated CW mode is given by $|\delta\tilde{a}^{(0)}_{+}|^2$. This means that as long as backscattering couples light from the CCW into the CW mode (i.e. for $\tilde{\beta}_{\pm} \neq 0$) the CW mode will host a finite emission. Note that the intensity in the CW mode increases whenever the system features a smaller value of $g$, that is a more nonlocal gain saturation. The perturbation theory breaks down for a pure thermo-optically driven gain saturation as Eq.~\eqref{eq:PertTheory} diverges for $g=1$. This is a consequence of the threefold-degenerate Goldstone mode that appears in this case.

Our simulations confirm that Eq.~\eqref{eq:PertTheory} correctly accounts for the intensity of the perturbatively populated CW mode as long as $|\tilde{\beta}_{\pm}|\ll \gamma_{\rm T}$. The intensity in the preferred mode is still given with a large precision by the unperturbed solution $|\tilde{a}^{(0)}_{-,\rm un}|^2$ and the eigenvalues of the matrix $A$ in Eq.~\eqref{eq:DiffEqsMatrix} do not significantly change. Therefore we conclude that in the presence of perturbative backscattering both the ring resonator and the TJR lase unidirectionally to a great extent, although some light is also present in the unfavored mode at the same frequency, with an intensity proportional to the backscattering coupling in its direction.


\subsection{Ring resonator}
\label{sec:MicBS_RingResonator}

Here we demonstrate how the values of the Hermitian and non-Hermitian coefficients describing backscattering~(\ref{eq:n}-\ref{eq:h}) control the lasing chirality in a ring resonator. This fact can be employed to construct unidirectional lasers by properly engineering the resonator's microscopic backscattering, for example by means of one or more nanotips coupled with the evanescent field of the resonator, similarly to the experiment of~\cite{Peng_2016}. Note that in a ring resonator $\gamma_{\rm S}=0$ and therefore $\gamma_{\rm T}=\gamma_{\rm A}$.

As was shown in Sec.~\ref{sec:PerturbativeSolution} the presence of perturbative backscattering does not appreciably change the fluctuation dynamics eigenvalues and the system lases unidirectionally to a great extent. Nevertheless, for values $|\tilde{\beta}_{\pm,\mp}| \gtrsim 10^{-2} |\gamma_{\rm A}|$ the perturbation theory~\eqref{eq:PertTheory} breaks down, a significant intensity populates the unfavored lasing direction and the intensity in the preferred direction falls below its usual value $P_{0}/\gamma_{\rm A}-1$. Fig.~\ref{fig:SmallBS} shows the intensities (panel (a)) and real part of the eigenvalues of matrix $A$ (panel (b)) for a ring resonator featuring $g=2$, $h=0$ and $|n|=0.1\gamma_{\rm A}$. 
Above the lasing threshold located at $P_{0} \simeq 0.9\gamma_{\rm A}$ both counterpropagating modes are amplified with the same intensity and oscillate simultaneously in a coherent way. As $P_{0}$ and $|\tilde{a}^{(0)}_{\pm}|^2$ grow the effect of mode competition in the gain saturation gets reinforced and intensity fluctuations become more susceptible of breaking the symmetry of the laser emission. This is evident from panel (b), where one of the real parts quickly grows as $P_{0}$ increases and at some point turns positive. Here, the system undergoes a transition towards a regime with a larger lasing intensity in a preferred mode. The larger $g$ (that is, the larger the local character of the gain saturation), the smaller the value of $P_{0}$ necessary to drive the system out of the bidirectional emission state as gain saturation asymmetry becomes more important. Nevertheless, the unfavored mode maintains a small intensity due to the finite backscattering in both directions.

The smaller lasing threshold obtained in Fig.~\ref{fig:SmallBS} can be understood by diagonalizing the matrix $A$ of Eq.~\eqref{eq:DiffEqsMatrix}. In the general coupling case $\beta_{\pm,\mp}\neq 0$ below the lasing threshold (i.e. for $|\tilde{a}^{(0)}_{\pm}|^2=0$) the eigenvalues are
\begin{align}
    \lambda_{1}=+(\omega_{0}-\omega)+i(P_{0}-\gamma_{\rm A})+(\beta^{}_{\pm}\beta^{}_{\mp})^{1/2},\\
    \lambda_{2}=-(\omega_{0}-\omega)+i(P_{0}-\gamma_{\rm A})-(\beta^{*}_{\pm}\beta^{*}_{\mp})^{1/2},\\
    \lambda_{3}=+(\omega_{0}-\omega)+i(P_{0}-\gamma_{\rm A})-(\beta^{}_{\pm}\beta^{}_{\mp})^{1/2},\\
    \lambda_{4}=-(\omega_{0}-\omega)+i(P_{0}-\gamma_{\rm A})+(\beta^{*}_{\pm}\beta^{*}_{\mp})^{1/2}.
\end{align}
The terms in the square roots can modify their real parts, therefore shifting the threshold position w.r.t. the backscattering-free case, where these terms were zero. The new lasing threshold takes place at a power $P_{0}=\gamma_{\rm A}-Im\{(\beta_{\pm}\beta_{\mp})^{1/2}\}$. If $n=0$ one has that $\beta_{\mp}=\beta^{*}_{\pm}$ and the square roots are purely real. In this case the lasing threshold remains at its usual position $P_{0}=\gamma_{\rm A}$. The same is true for a unidirectional coupling (featuring either $\beta_{\pm}=0$ or $\beta_{\mp}=0$). The maximum shift for fixed $|\beta_{\pm,\mp}|$ occurs for $h=0$ and is given by $P_{0}=\gamma_{\rm A}-|\beta_{\pm}|$.

In spite of the finite intensity emitted in the unfavored direction, in the small backscattering regime the majority of the ring resonator's emission still takes place in a preferential direction determined by the particular choice of backscattering coefficients. Fig.~\ref{fig:BackscatteringRing} shows the probability of a ring resonator featuring $g>1$ to lase preferentially in a certain direction as a function of the Hermitian and non-Hermitian parameters. Here, the probability is calculated by averaging over many independent realizations of the initial noise used to seed the laser operation.

We first study the situation in which both of them are finite and of equal strength, i.e. $|h|=|n|\neq 0$. In panel (a) these are kept at a fixed modulus $|h|=|n| \simeq 5\times 10^{-4}\gamma_{\rm A}$ while the phase angle between them $\Delta\phi=\phi_{n}-\phi_{h}$ is varied. For $\Delta\phi=0,\pi$ (which implies $|\beta_{\pm}|=|\beta_{\mp}|$) the system has equal probabilities of lasing in each mode, while for $\Delta\phi=\pm\pi/2$ the emission preferentially takes place in one particular direction with 100$\%$ probability. It is easy to realize that $\Delta\phi=+\pi/2$ implies $\beta_{\mp}=0$ while $\Delta\phi=-\pi/2$ corresponds to having $\beta_{\pm}=0$. These results are in perfect agreement with the behavior of passive microdisk resonators reported in~\cite{Biasi_2019}. 

However, this is only valid when $|h|=|n|\neq 0$. If one of the two parameters is kept fixed and the other is reduced to zero, the lasing probability in each direction tends to $\mathcal{P}=0.5$ regardless of the phase difference $\Delta\phi$, as shown in panels (b) and (c). This is easily understood by exploring Eqs.~(\ref{eq:n}-\ref{eq:h}): either $n=0$ or $h=0$ imply in fact an equal coupling strength in the two directions, i.e. $|\beta_{\pm}|=|\beta_{\mp}|$.

\begin{figure}
    \centering
    \includegraphics[width=0.5\textwidth]{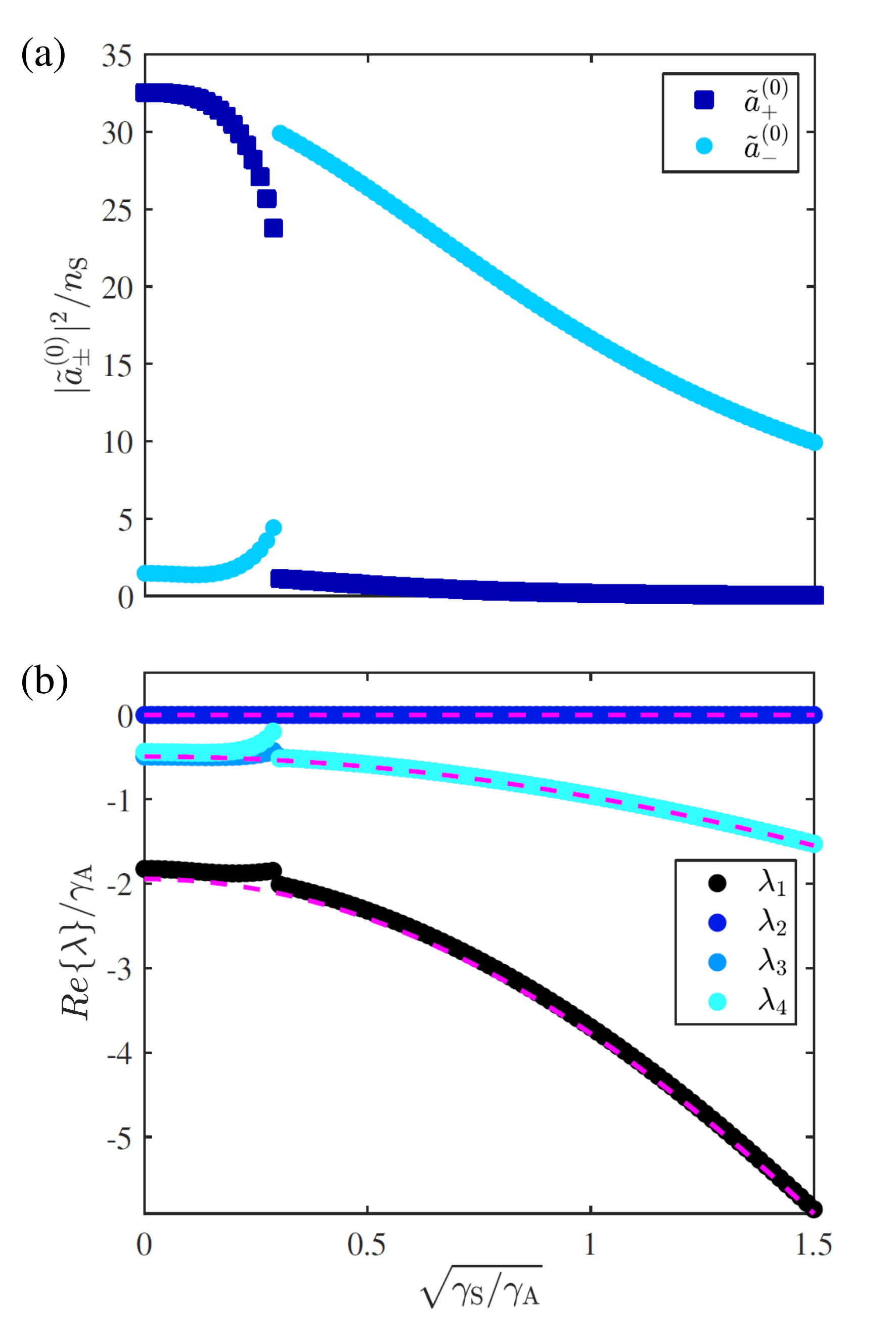}
    \caption{Ramp in the S-waveguide coupling from $\gamma_{\rm S}=0$ to $\sqrt{\gamma_{\rm S}/\gamma_{\rm A}}=1.5$ for a ring resonator with $g=2$ in the presence of a backscattering described by $|n|=0.1\gamma_{\rm A}$ and $h=0$ for a fixed pump rate $P\simeq 36\gamma_{\rm A}$. \textbf{(a)} Intensities in each mode $|\tilde{a}^{(0)}_{\pm}|^2$ as a function of the square root of the S-coupling losses $\sqrt{\gamma_{\rm S}}$. \textbf{(b)} Real part of the linearized fluctuation dynamics eigenvalues $\lambda$ as a function of $\sqrt{\gamma_{\rm S}}$. The dashed lines correspond to the backscattering-free eigenvalues~(\ref{eq:EigenvaluesAboveThreshold1}-\ref{eq:EigenvaluesAboveThreshold4}). Where not visible, $Re\{\lambda_{3}\}$ lies below $Re\{\lambda_{4}\}$.}
    \label{fig:SmallBS_TurningOnTheS}
\end{figure}
%


\subsection{TJR}
\label{sec:MicBS_TJR}

In the case of a TJR coupling unidirectionally the CW into the CCW direction one has that $|\beta_{\mp}|\gg|\beta_{\pm}|$ and therefore
\begin{align}
    h\simeq & -\frac{\beta_{\mp}}{2},\\
    n\simeq &\;\; i \; \frac{\beta_{\mp}}{2},
\end{align}
which falls into the $|n|=|h|$, $\Delta\phi=-\pi/2$ case. As already shown in this Section, this implies that even though the resonator hosts a small intensity in the CW mode due to the finite coupling $\beta_{\pm}\neq 0$, the majority of the system emission takes place in the CCW mode. The presence of additional backscattering coupling light into the CCW direction does not have any visible effect as it does not modify the coupling $\beta_{\mp}$ significatively~\footnote{For a TJR with an oppositely oriented S-element, one still has $|n|=|h|$ but $\Delta\phi=\pi/2$. As expected, this implies a preferred emission in the CW direction.}.

\begin{figure*}
    \centering
    \includegraphics[width=\textwidth]{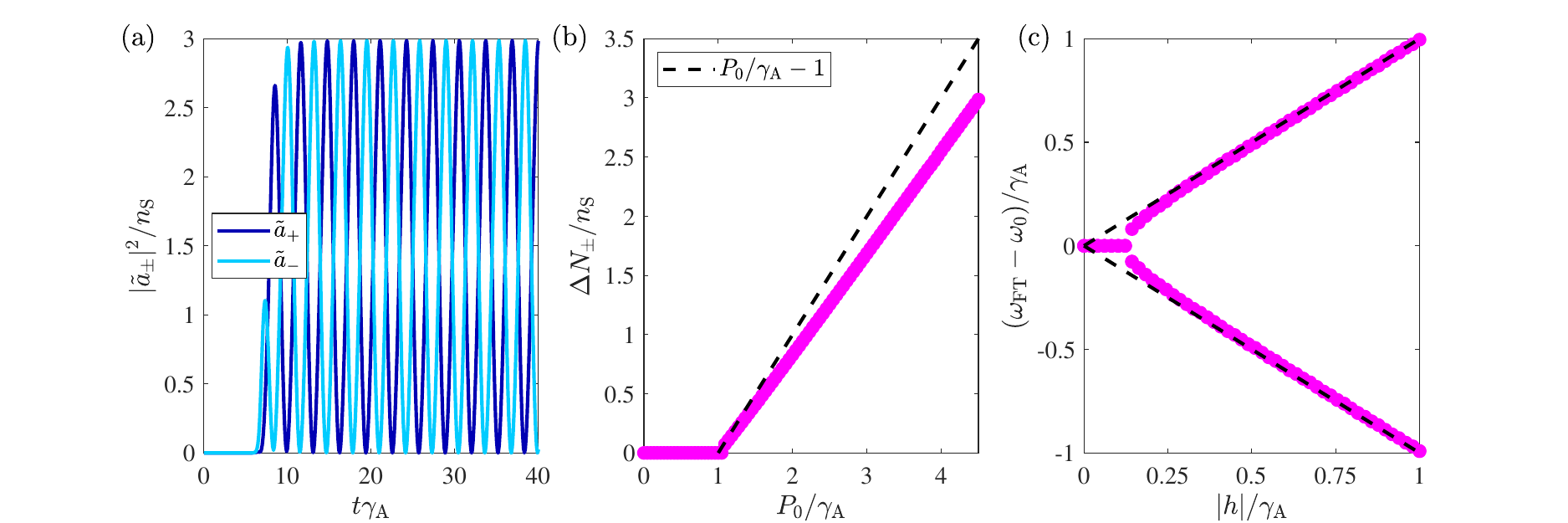}
    \caption{Ring resonator with g= 2 and backscattering given by $|h|= \gamma_{\rm A}$ and $n=0$. \textbf{(a)} Intensity in each mode $|\tilde{a}_{\pm}|^2$ as a function of time $t$. The pump rate is fixed at $P_{0}=4.5\gamma_{\rm A}$. \textbf{(b)} Amplitude $\Delta N_{\pm}$ of the oscillations of the intensity in each mode as a function of the pump rate $P_{0}$ (pink circles). The amplitude is the same in the two directions. The dashed line corresponds to the usual dependence in the absence of backscattering. \textbf{(c)} Frequencies $\omega_{\rm FT}$ of the two emission components (whose beating leads to the intensity oscillations visible in panel \textbf{(a)}) as a function of the Hermitian backscattering coefficient $|h|$. The pump rate is fixed at $P_{0}=4.5\gamma_{\rm A}$. The frequencies are the same in the two directions.}
    \label{fig:LargeBS}
\end{figure*}

In order to shine more light into the role of the S waveguide we performed a coupling ramp from $\gamma_{\rm S}=0$ (which corresponds to a ring resonator) to $\sqrt{\gamma_{\rm S}/\gamma_{\rm A}}=1.5$ in a resonator featuring $g=2$ and a fixed pump rate $P_{0} \simeq 36 \gamma_{\rm A}$. The initial situation is described by the backscattering parameters employed in Fig.~\ref{fig:SmallBS}, namely $|n|=0.1 \gamma_A$ and $h=0$, for which the two directions have equal probabilities of hosting the preferential lasing emission. Our results are displayed in Fig.~\ref{fig:SmallBS_TurningOnTheS}. As $\gamma_{\rm S}$ grows $\beta_{\mp}$ increases: This leads to an intensity transfer from the CW to the CCW mode. For $\sqrt{\gamma_{\rm S}/\gamma_{\rm A}}\gtrsim 0.25$ the S-waveguide coupling strength grows beyond the backscattering couplings and the real part of one of the eigenvalues of matrix~\eqref{eq:DiffEqsMatrix} crosses zero from below. The system then experiences a transition towards a state with preferential lasing in the CCW direction. As $\gamma_{\rm S}$ continues growing the relative importance of backscattering w.r.t. the S coupling decreases. For $\sqrt{\gamma_{\rm S}/\gamma_{\rm A}}\simeq 1$ the ratio between backscattering in the CW direction $\beta_{\pm}$ and the total loss rate $\gamma_{\rm T}$ already gives $|\beta_{\pm}| \simeq 5\times 10^{-2}\gamma_{\rm T}$, which means that the system is approaching the perturbative backscattering regime described in Sec.~\ref{sec:PerturbativeSolution}. As this happens the spectrum of eigenvalues of the matrix $A$ in Eq.~\eqref{eq:DiffEqsMatrix} approaches the backscattering-free eigenvalues~(\ref{eq:EigenvaluesAboveThreshold1}-\ref{eq:EigenvaluesAboveThreshold4}) and the intensity ratio between the two directions already gives a sizable value $|\tilde{a}^{(0)}_{-}|^2 \simeq 250 |\tilde{a}^{(0)}_{+}|^2$ for $\sqrt{\gamma_{\rm S}/\gamma_{\rm A}}=1.5$. The eventual decrease of the intensity also in the CCW direction that is observed at higher $\gamma_{\rm S}$ is due to the growth of the total loss rate $\gamma_{\rm T}$ that is naturally associated to the increasing $\gamma_{\rm S}$.

These results confirm the possibility to use the S waveguide in order to guarantee unidirectional lasing. The necessary condition is to implement a sufficiently large coupling allowing to treat backscattering as a microscopic perturbation to unidirectional lasing, which according to our model is legitimate at least up to $|\tilde{\beta}_{\pm,\mp}| \lesssim 10^{-2} \gamma_{\rm T}$.


\section{Large backscattering}
\label{sec:LargeBackscattering}

In this Section we investigate lasing in ring resonators and TJRs characterized by the presence of a large Hermitian backscattering compared with the resonator loss rate in the absence of the S element, i.e. $|h|\gtrsim\gamma_{\rm A}$. As we will show such a coupling introduces self-oscillations of the intensity in the two directions. The requirement $|n|\leq\gamma_{\rm T}$ that is needed to avoid an unphysical backscattering-induced gain implies that our model can only account for a large Hermitian backscattering. For simplicity, we will then consider $n=0$ throughout this Section. The addition of a non-Hermitian coupling would only introduce an asymmetry between the two counter-propagating directions that damps the oscillating behavior for values of $n$ approaching $\gamma_{\rm T}$. As in Sec.~\ref{sec:MicBS}, the analysis carried in this Section does not take into account the resonance shift due to the nonlinearity, i.e. $n_{\rm NL}=0$. This additional effect will be studied in Sec.~\ref{sec:Nonlinear}. Under this condition, lasing occurs at the resonator frequency $\omega_0$.

Fig.~\ref{fig:LargeBS}a shows the time evolution of the intensity in each mode for a ring resonator featuring $g=2$, $n=0$, and $|h|=\gamma_{\rm A}$ at a pump rate $P_{0}=4.5\gamma_{\rm A}$. As observed in previous works~\cite{Schwartz_2007,Ceppe_2019} for such a large backscattering coupling the resonator enters a regime in which the intensity in the two directions oscillates in phase opposition at a frequency given by twice the modulus of the backscattering coupling $|h|$. Panel (b) shows the amplitude $\Delta N_{\pm}$ of these oscillations as a function of $P_{0}$. This is equal in the two directions and is slightly smaller than the backscattering-free amplitude, which is given by $P_{0}/\gamma_{\rm A}-1$. The frequency at which the field amplitudes $\tilde{a}_{\pm}$ oscillate $\omega_{\rm FT}$ is extracted from a Fourier transform, and displayed in panel (c) for a fixed pump rate $P_{0}=4.5\gamma_{\rm T}$ as a function of $|h|$. As shown in Sec.~\ref{sec:MicBS} for a small backscattering in our reference frame rotating at a frequency $\omega=\omega_{0}$ the field amplitudes do not oscillate. As $|h|$ grows beyond approximately $0.1\gamma_{\rm A}$ both field amplitudes start oscillating with two opposite frequency components which rapidly approach $\pm |h|$ as the Hermitian backscattering grows.

\begin{figure}
    \centering
    \includegraphics[width=0.5\textwidth]{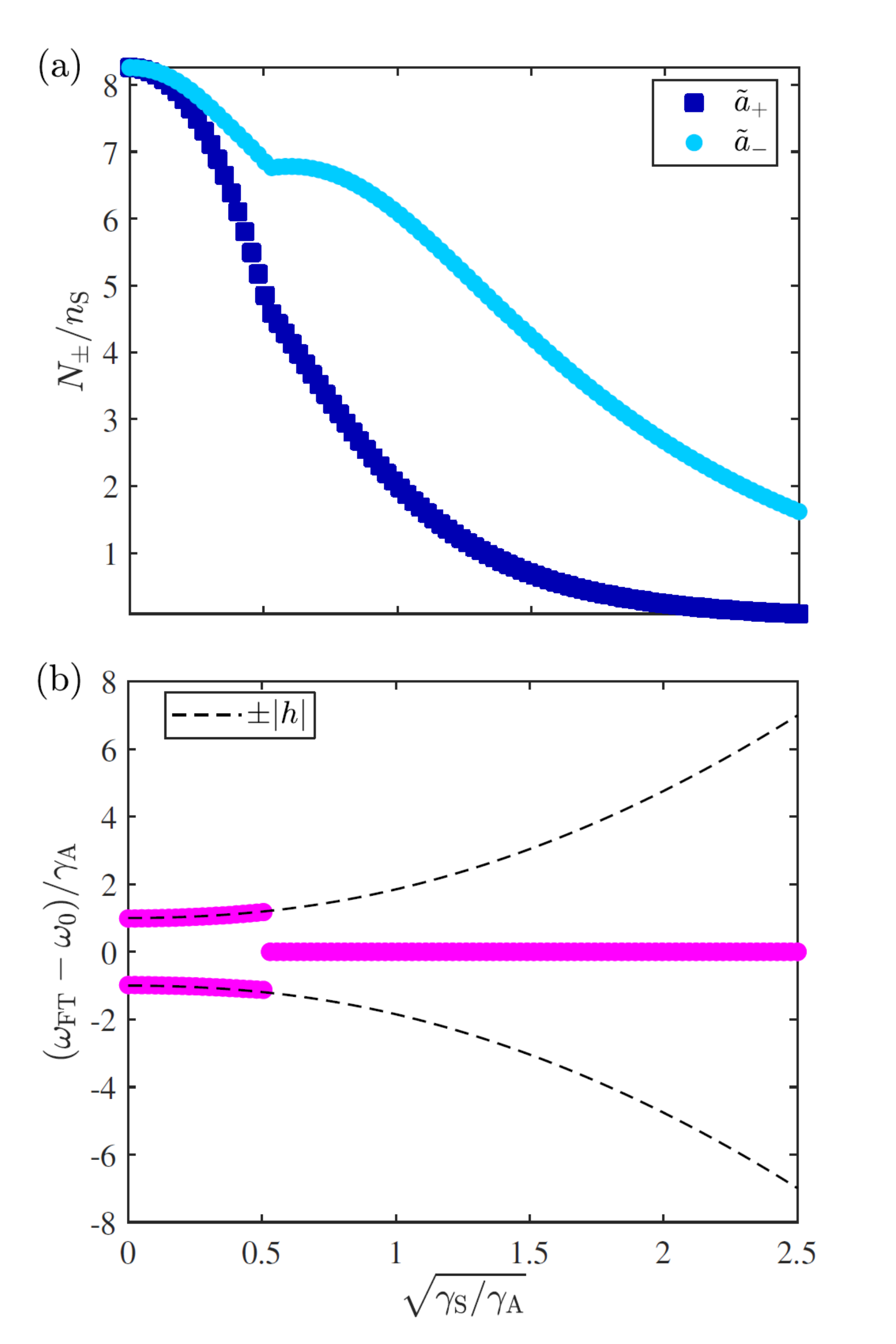}
    \caption{Ramp in S-waveguide coupling from $\gamma_{\rm S}=0$ (ring resonator) to $\sqrt{\gamma_{\rm S}/\gamma_{\rm A}}=2.5$ for a ring resonator with $g=2$ in the presence of a backscattering given by $n=0$ and $|h|=\gamma_{\rm A}$ for a fixed pump rate $P\simeq 20\gamma_{\rm A}$. \textbf{(a)} Mean value between intensity maxima and minima in each direction $N_{\pm}$ as a function of the square root of the S-coupling losses $\sqrt{\gamma_{\rm S}}$. Data corresponding to $\sqrt{\gamma_{\rm S}/\gamma_{\rm A}}\lesssim 0.5$ falls into the oscillation regime. \textbf{(b)} Frequency of the field amplitudes in the two directions $\omega_{\rm FT}$ as a function of $\sqrt{\gamma_{\rm S}}$.}
    \label{fig:LargeBS_TurningOnTheS}
\end{figure}

This situation changes dramatically when the S-shaped waveguide of the TJR is introduced. In Fig.~\ref{fig:LargeBS_TurningOnTheS} a ring resonator featuring the same $g$ parameter and backscattering coefficients as in the simulation displayed in Fig.~\ref{fig:LargeBS} is subjected to an S-coupling ramp. The pump rate is fixed at $P_{0}=20\gamma_{\rm A}$. Panel (a) shows the mean value between the maximum and minimum intensities $N_{\pm}=(max\{|\tilde{a}_{\pm}|^2\}+min\{|\tilde{a}_{\pm}|^2\})/2$ emitted in each direction. In the oscillatory regime that takes place for $\sqrt{\gamma_{\rm S}/\gamma_{\rm A}}\lesssim 0.5$ this quantity corresponds to half the amplitude of the oscillations. In the nonoscillatory regime (for $\sqrt{\gamma_{\rm S}/\gamma_{\rm A}}\gtrsim 0.5$) one has that $max\{|\tilde{a}_{\pm}|^2\}=min\{|\tilde{a}_{\pm}|^2\}$ and therefore $N_{\pm}=|\tilde{a}^{(0)}_{\pm}|^2$. On the other hand, Fig.~\ref{fig:LargeBS_TurningOnTheS}b shows the oscillation frequency $\omega_{\rm FT}$ extracted from a Fourier transform of the field amplitudes $\tilde{a}_{\pm}$ in each direction for the same ramp in $\gamma_{\rm S}$. In the oscillatory regime the two amplitudes oscillate with two frequency components given by the Hermitian backscattering coefficient $\pm |h|$. 
For $\sqrt{\gamma_{\rm S}/\gamma_{\rm A}}\gtrsim 0.5 $ the coupling strength with the S waveguide increases beyond the backscattering couplings given by $|h|=\gamma_{\rm A}$ and the oscillations disappear (see panel (b)). This regime is equivalent to the situation studied in Sec.~\ref{sec:MicBS_TJR}. Panel (a) shows that the S-shaped waveguide imposes a definite chirality in the laser emission, as the CCW mode becomes the favored mode when $\gamma_{\rm S}$ is increased. Once again, the eventual decrease of the intensity in the two directions that is visible at larger $\gamma_{\rm S}$ is due the growth of the total loss rate $\gamma_{\rm T}$. Interestingly, this effect mainly affects the unfavored CW mode, whose emission becomes rapidly negligible. For instance, already at $\sqrt{\gamma_{\rm S}/\gamma_{\rm A}}=2.5$ the intensity ratio between both directions is $|\tilde{a}^{(0)}_{-}|^2\simeq 17 |\tilde{a}^{(0)}_{+}|^2$.

\begin{figure}[t]
    \centering
    \includegraphics[width=0.5\textwidth]{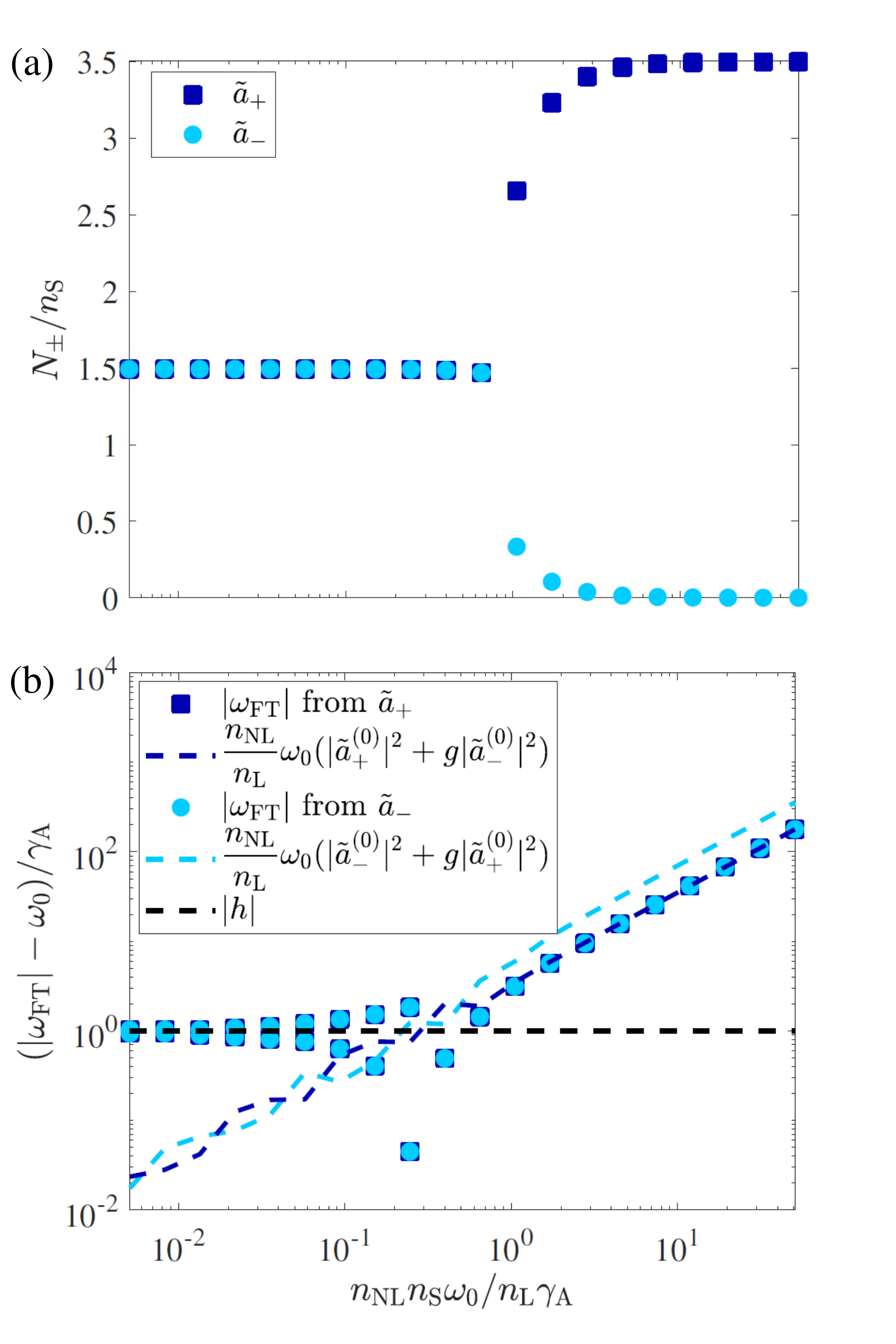}
    \caption{\textbf{(a)} Mean value between intensity maxima and minima $N_{\pm}$ as a function of the nonlinear refractive index $n_{\rm NL}$ for a ring resonator featuring a Kerr-like nonlinearity ($g=2$) and backscattering given by $n=0$, $|h|=\gamma_{\rm A}$. The pump rate is fixed at $P_{0}=4.5\gamma_{\rm A}$. Data corresponding to $n_{\rm NL}n_{\rm S}\omega_{0}/n_{\rm L}\gamma_{\rm A}\lesssim 1$ fall into the oscillatory regime. \textbf{(b)} Absolute values of the oscillation frequencies $\omega_{\rm FT}$ of the field amplitudes $\tilde{a}_{\pm}$ as a function of $n_{\rm NL}$. Squares and circles are the values extracted from the Fourier transform (FT) of the data presented in panel (a). Dashed lines represent the expected oscillation frequency in the linear regime $|h|$ and the nonlinear frequency shifts for each mode.}
    \label{fig:TurnOnNonlinearity_LargeBS}
\end{figure}
\begin{figure}[t]
    \centering
    \includegraphics[width=0.5\textwidth]{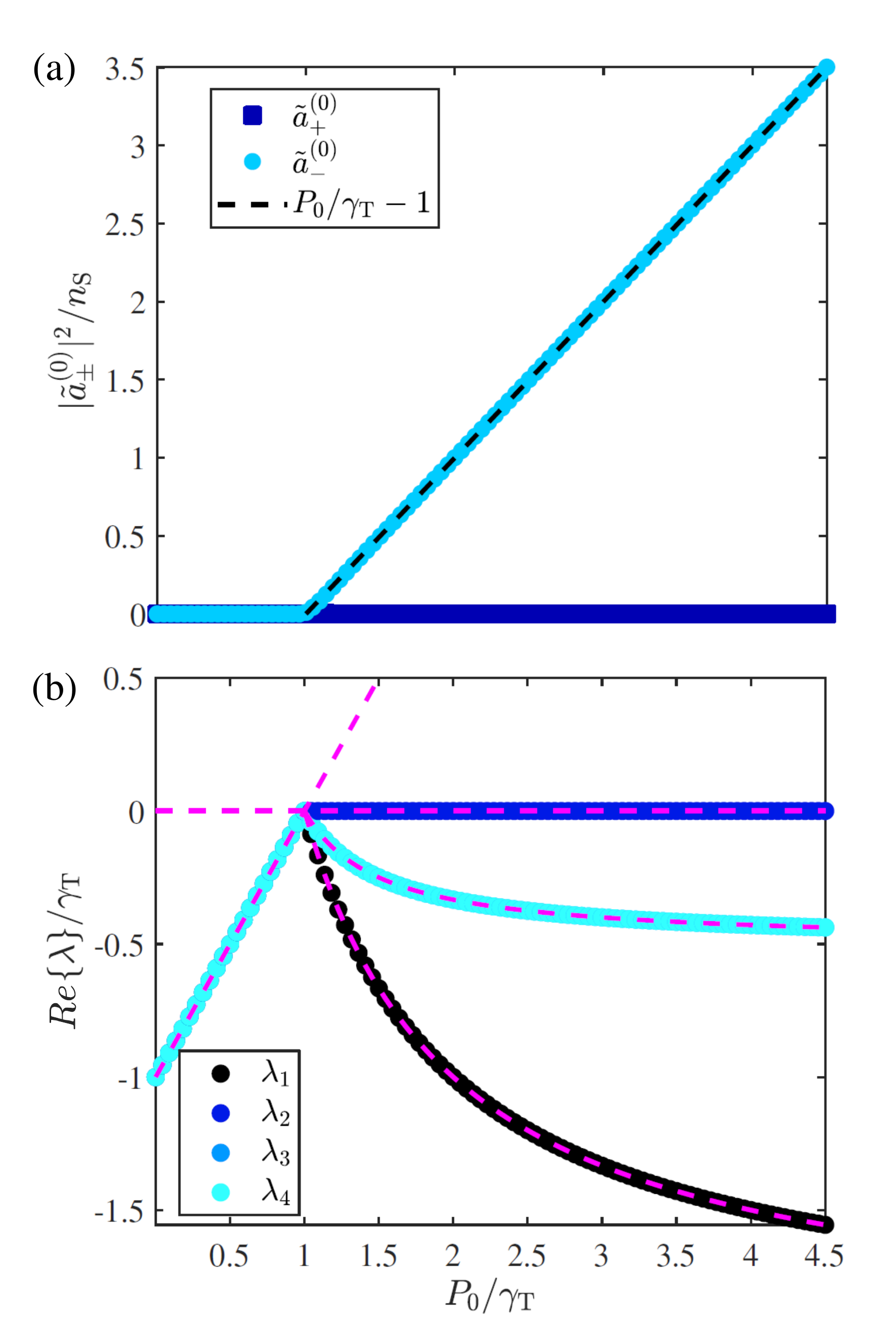}
    \caption{Pump rate ramp in a TJR featuring an S-waveguide coupling $\sqrt{\gamma_{\rm S}/\gamma_{\rm A}}=2.5$, a $g=2$ nonlinearity given by $n_{\rm NL}n_{\rm S}\omega_{0}/n_{\rm L}\gamma_{\rm A}=50$ and backscattering parameters $n=0$ and $|h|=\gamma_{\rm A}$. \textbf{(a)} Steady-state intensity $|\tilde{a}^{(0)}_{\pm}|^2$ in each mode as a function of the pump rate $P_{0}$. The dashed line is the backscattering-free intensity. \textbf{(b)} Real part of the fluctuation dynamics eigenvalues $\lambda$ as a function of $P_{0}$. The dashed lines correspond to the real parts of the backscattering-free eigenvalues given by Eqs.~(\ref{eq:EigenvaluesBelowThreshold12}-\ref{eq:EigenvaluesAboveThreshold4}). $Re\{\lambda_{3}\}$ lies below $Re\{\lambda_{4}\}$. Below threshold, the real parts are degenerate.}
    \label{fig:NonlinearTaiji}
\end{figure}
%


\section{Effect of the nonlinearities}
\label{sec:Nonlinear}

In this Section we show that the local Kerr nonlinearity indigenous to the waveguide material reinforces the unidirectionality of the ring laser emission. This effect can be combined with the action of the S-waveguide in order to further reinforce unidirectional lasing even in the large backscattering regime. The nonlinearity modifies the resonance frequency of each mode $\omega^{(0)}_{\pm}$ according to
\begin{equation}
    \omega^{(0)}_{\pm}=\omega_{0}\left[1-\frac{n_{\rm NL}}{n_{\rm L}}(|\tilde{a}^{(0)}_{\pm}|^2+g|\tilde{a}^{(0)}_{\mp}|^2)\right].
\label{eq:NonlinearResonanceFreq}
\end{equation}

In the backscattering-free ring resonator and unidirectional TJR lasers the nonlinearity has no effects beyond the shift of the resonance frequency. The behavior is therefore the same as the one described in Sec.~\ref{sec:BackscatteringFreeResonators} for the linear regime. The same is true for a pure thermo-optic nonlinearity with $g=1$ even in the presence of backscattering. In this case the resonance frequency of the two modes is in fact shifted by equal amounts and therefore the coupling between them is not perturbed by the nonlinearity. On the other hand, if one has $g>1$ and finite and unequal intensities in the two modes, as is possible for a sufficiently large backscattering in a ring resonator, the resonance frequency in the two directions will be different. This fact further suppresses the intermodal coupling as it reduces the probability of light to scatter from one mode to another.

Fig.~\ref{fig:TurnOnNonlinearity_LargeBS}a shows the mean value between the intensity maxima and minima $N_{\pm}$ as the nonlinear refractive index $n_{\rm NL}$ is varied. In the nonoscillatory regime this quantity reduces to the steady-state intensity $|\tilde{a}^{(0)}_{\pm}|^2$. Panel (b) of the same figure shows the field oscillation frequencies $\omega_{\rm FT}$ in the two directions for the same $n_{\rm NL}$ ramp, as extracted from a Fourier transform of the field amplitudes. Once again we chose a rotating reference frame at the linear resonance frequency $\omega_{0}$. The calculation was made for a ring resonator featuring a $g=2$ nonlinearity and the backscattering parameters $n=0$, $|h|=\gamma_{\rm A}$, which fall into the large backscattering regime described in Sec.~\ref{sec:LargeBackscattering}. The pump rate is fixed at $P_{0}=4.5\gamma_{\rm A}$.  Similarly to Fig.~\ref{fig:LargeBS_TurningOnTheS}, as $n_{\rm NL}$ increases oscillations disappear, leading to a regime in which the laser emission is concentrated in a randomly chosen direction (the CW one in the figure), each with 50$\%$ of probability. This effect is reinforced as $n_{\rm NL}$ grows, ultimately achieving a pure unidirectional emission for $n_{\rm NL}n_{\rm S}\omega_{0}/n_{\rm L}\gamma_{\rm A}\gtrsim 10$. For the largest value of $n_{\rm NL}$ calculated, the intensity ratio gives $|\tilde{a}^{(0)}_{+}|^2 \simeq 10^{4}|\tilde{a}^{(0)}_{-}|^2$. 

As shown in panel (b), for negligible values of $n_{\rm NL}$ both modes oscillate with two frequency components given by the Hermitian backscattering coefficient $\pm |h|$, as demonstrated in Sec.~\ref{sec:LargeBackscattering}. For the purpose of using a logarithmic scale, the absolute values $|\omega_{\rm FT}|$ are displayed. As $n_{\rm NL}$ increases, the frequencies blue shift until a transition takes place at $n_{\rm NL}n_{\rm S}\omega_{0}/n_{\rm L}\gamma_{\rm A}\gtrsim 1$ back to a regime with stationary intensity values, in which the two field amplitudes oscillate at a single frequency given by the nonlinear displacement of the resonance frequency of the resonator (Eq.~\eqref{eq:NonlinearResonanceFreq}) for the preferred lasing direction. This is telling us that the unfavored CCW direction does no longer feature laser oscillations at its own resonance frequency but that all the light that populates it is being backscattered from the preferential CW direction.

Finally, Fig.~\ref{fig:NonlinearTaiji} displays the steady-state intensities $|\tilde{a}^{(0)}_{\pm}|^2$ and the corresponding real part of the eigenvalues of matrix $A$ (Eq.~\eqref{eq:DiffEqsMatrix}) for a pump rate ramp in a TJR featuring an S-waveguide coupling $\sqrt{\gamma_{\rm S}/\gamma_{\rm A}}=2.5$, a $g=2$ Kerr nonlinearity of strength $n_{\rm NL}n_{\rm S}\omega_{0}/n_{\rm L}\gamma_{\rm A}=50$, and large backscattering parameters $n=0$ and $|h|=\gamma_{\rm A}$. In the absence of the S-waveguide and the nonlinearity, the corresponding ring resonator would feature intensity oscillations at a frequency given by $2|h|$. Instead, the nonlinear TJR capitalizing on these two crucial elements shows unidirectional lasing with a preferred CCW chirality with 100$\%$ probability. The intensity in the CW direction is four orders of magnitude smaller than that in the CCW one and can therefore be safely neglected. The Bogoliubov analysis of the small perturbations to this steady state reveals that the eigenvalues of the linearized fluctuation dynamics matrix $A$ are identical to those calculated in the backscattering-free case, ensuring the stability of the unidirectional emission.

%

\section{Conclusions}
\label{sec:Conclusions}

In this work we demonstrated that an active “Taiji” micro-ring resonator (TJR) formed by a standard ring resonator supplemented by an S-shaped element unidirectionally coupling the two counterpropagating modes shows a preferential chirality in the laser oscillation even in the presence of a large backscattering. The presence of the S-shaped element implies robust unidirectional lasing in the favored direction and restricts the emission in the other direction to negligible values. 

Our theoretical investigation is based on the most general version of the coupled-mode equations of motion for the field amplitudes in the two counter-propagating modes. These equations are first solved for the steady-states. The dynamical stability of these latter against small perturbations is then assessed within a linearized theory by looking at the real part of the eigenvalues of the linearized fluctuations dynamics matrix.

In the absence of backscattering, one of the two counterpropagating lasing solutions of the ring resonator disappears for a sufficiently strong coupling by the S-element, leaving a single stable solution with laser emission only in the mode which is favored by the S-shaped element. This unidirectional emission remains stable even in the presence of backscattering effects due, e.g., to the roughness of the resonator surface, as long as the coupling by the S-element exceeds the backscattering strength. The robustness of the unidirectional laser emission is further reinforced by a Kerr nonlinearity shifting the resonance of the two counterpropagating modes by different amounts.

While unidirectional laser emission in a random direction in ring resonators can be seen as a spontaneous breaking of the time-reversal $\mathcal{T}$-symmetry in the lasing state above threshold, the explicit breaking of $\mathcal{P}$-symmetry in the geometrical shape of our Taiji resonator leads to a preferred chirality of the laser emission. This can be understood as an explicit dynamical breaking of $\mathcal{T}$-symmetry induced by the broken $\mathcal{P}$-symmetry in an otherwise $\mathcal{T}$-symmetric device. 

This novel mechanism for breaking $\mathcal{T}$-reversal appears of great interest in view of realizing optical isolators and other non-reciprocal devices based on wave-mixing phenomena without the need for magnetic elements. In the context of topological photonics~\cite{Ozawa_2019,Bandres_2018}, this will allow to dynamically generate a topological Chern insulator using structures based on non-magnetic active dielectric materials. These exciting topics are the subject of on-going work. 


%
\acknowledgements
We acknowledge financial support from the European Union FET-Open grant ``MIR-BOSE'' (n. 737017), from the H2020-FETFLAG-2018-2020 project ``PhoQuS'' (n.820392), from the Provincia Autonoma di Trento, from the Q@TN initiative, and from Google via the quantum NISQ award. Stimulating discussions with Stefan Rotter, Yannick Dumeige, Sylvain Schwartz, Petr Zapletal, Crist\'obal Lled\'o, Marzena Szyma\'nska, Alberto Amo, and Jacqueline Bloch are warmly acknowledged.


\bibliography{SingleLasingTaiji.bib}

\end{document}